\documentclass[prd,aps,superscriptaddress,nofootinbib,amsmath,amssymb,showpacs]{revtex4}
\usepackage{graphicx}
\usepackage{dcolumn}
\usepackage{bm}
\usepackage{multirow}
\usepackage{adjustbox}
\usepackage{textcomp}
\usepackage{mathtext}
\usepackage{hyperref}
\graphicspath{{./eps/}}

\begin{document}

\title{{\Large Quasielastic production of polarized $\tau$ leptons in $\nu_{\tau}$ and $\bar{\nu}_{\tau}$ scattering from nucleons}}

\author{A. \surname{Fatima}}
\affiliation{Department of Physics, Aligarh Muslim University, Aligarh-202002, India}
\author{M. Sajjad \surname{Athar}}
\email{sajathar@gmail.com}
\affiliation{Department of Physics, Aligarh Muslim University, Aligarh-202002, India}
\author{S. K. \surname{Singh}}
\affiliation{Department of Physics, Aligarh Muslim University, Aligarh-202002, India}
 
\begin{abstract}
 The cross sections and polarization components of the $\tau$ leptons produced in the charged current induced quasielastic $\nu_\tau~(\bar\nu_\tau) - N$ scattering have been studied. The theoretical 
 uncertainties arising due to the use of different vector form factors and the axial dipole mass in the axial vector form factor have been investigated. Due to the high mass of $\tau$ lepton, the contributions from the term containing pseudoscalar and second class current form factors are non-negligible and contribute to the uncertainty in the cross section and polarization observables as these form factors are not well known. 
 In view of the currently proposed experiments by DUNE, SHiP and DsTau collaborations to study the production of $\tau$ lepton, an updated calculation of the cross sections and polarizations of tau leptons in the case of quasielastic production have been done and the numerical results have been presented along with a discussion of the theoretical uncertainties. 
 \end{abstract}
\pacs{{12.15.-y}, {13.88.+e}} 
\maketitle

 \section{Introduction}
 The production of $\tau$ leptons in the $\nu_{\tau}~(\bar{\nu}_{\tau})-$nucleon interactions induced by the charged current is of considerable topical interest in the study of neutrino oscillations. The observation of a few $\tau$ lepton events in the recent experiments by the DONUT~\cite{Kodama:2000mp, Kodama:2007aa} and the OPERA~\cite{Agafonova:2014bcr, Agafonova:2015jxn, Agafonova:2018auq} collaborations using accelerator neutrinos and by the Super-Kamiokande~\cite{Li:2017dbe, Abe:2012jj} and IceCUBE~\cite{Aartsen:2019tjl} collaborations using the atmospheric neutrinos have established the existence of $\nu_{\mu} \longrightarrow \nu_{\tau}$ oscillations. In view of the small number of $\tau$-lepton events in these experiments, new experiments have been proposed by the SHiP~\cite{Yoon:2020yrx, Anelli:2015pba, Alekhin:2015byh}, DsTau~\cite{Aoki:2019jry} and DUNE~\cite{Machado:2020yxl, Strait:2016mof, Abi:2018rgm} collaborations in which the number of lepton events produced in the $\nu_{\tau}-$nucleon interaction are expected to be of the order of a few thousands during the running time of 3--5 years. In the $\nu_{\tau}-$nucleon interactions, the tau~($\tau$) leptons are produced by the quasielastic~(QE), inelastic~(IE) and deep inelastic scattering~(DIS) processes. The relative importance of these processes depends upon the energy of the $\tau$ neutrinos~($\nu_{\tau}$). These $\tau$ neutrinos are produced either through the $\nu_{\mu} \longrightarrow \nu_{\tau}$ oscillations in the accelerator~\cite{Kodama:2000mp, Kodama:2007aa, Agafonova:2014bcr, Agafonova:2015jxn, Agafonova:2018auq, Machado:2020yxl, Strait:2016mof, Abi:2018rgm}, atmospheric~\cite{Li:2017dbe, Abe:2012jj, Aartsen:2019tjl} and astrophysical sources~\cite{Athar:1970esm} or through the decay of $D_{s}$ particles produced in the very high energy proton-hadron collisions~\cite{Yoon:2020yrx, Anelli:2015pba, Alekhin:2015byh}.
 
 The $\tau$ leptons produced in the $\nu_{\tau}-$nucleon interactions are identified by the leptons and pions into which these $\tau$ leptons decay through various leptonic and hadronic channels. The decay rates and the topologies of the leptons and pions produced in the decay processes of the $\tau$ lepton depend upon the production cross section and polarization of the $\tau$ leptons produced through the various reaction processes in the $\nu_{\tau}-$nucleon interactions. In the region of very high energy of $\nu_{\tau}$ where the lepton mass can be neglected $E_{\nu_{\tau}} \gg m_{\tau}$, the produced $\tau$ lepton is almost left handed and the cross section is dominated by the DIS processes. However, in the low and medium energy region in which the quasielastic and inelastic processes play a significant role, the knowledge of $\tau$ polarization, in addition to the cross section is an important parameter to determine the rate and the topological characteristics of the final leptons and pions coming from the $\tau$ decays, in order to reconstruct the genuine $\tau$ lepton events produced in the $\nu_{\tau}-$nucleon interactions. It is, therefore, important to study the theoretical uncertainties in the prediction of the cross section and the polarization of the $\tau$ produced in the 
 $\nu_{\tau}-$nucleon and also in the $\nu_{\tau}-$nucleus interactions as all the present and future experiments are using nuclear targets.
 
 In context of the present and future experiments which use or propose to use nuclear targets, the uncertainties involved in understanding the nuclear medium effects are an important source of systematic uncertainty in predicting cross sections and $\tau$ polarization. Moreover, there exist additional uncertainties even in the case of $\nu_{\tau} - N$ scattering as compared to $\nu_\mu - N$ scattering due to the large mass of $\tau$ lepton which makes the contribution of the pseudoscalar and second class current~(SCC) form factors significant~\cite{Athar:1970esm, Fatima:2018tzs}. Since the parametric form of the pseudoscalar form factor is not determined very well and is uncertain within the validity of partially conserved axial vector current~(PCAC) at high momentum transfers and the presence~(or absence) of the second class currents in the $\tau$ lepton sector is not known theoretically and/or phenomenologically, therefore, the contribution of these terms becomes model dependent leading to uncertainties in the cross sections and $\tau$ polarizations.
 
 There have been many calculations of the cross section and $\tau$ polarization induced by various processes in the $\nu_{\tau}-$nucleon scattering~\cite{Hagiwara:2003di, Hagiwara:2004gs, Hagiwara:2004xe, Aoki:2005wb, Kuzmin:2003ji, Kuzmin:2004ke, Graczyk:2004uy, Graczyk:2004vg, Graczyk:2017rti, Graczyk:2019blt, Graczyk:2019xwg, Conrad:2010mh, Bourrely:2004iy, Gazizov:2016dhn, Paschos:2001np, Lagoda:2007, Kurek:2005, Albright:1974ts}. While some of them consider explicitly the effect of nuclear medium in the quasielastic scattering~\cite{Graczyk:2004uy, Sobczyk:2019urm, Valverde:2006yi}, very few discusses the contribution of the pseudoscalar and/or second class current form factors~\cite{Hagiwara:2004gs, Kuzmin:2003ji, Kuzmin:2004ke, Graczyk:2004uy, Graczyk:2004vg}. None of them discuss the theoretical uncertainties due to the use of various parametric forms of the weak vector form factor arising specially due to the use of the charge form factor of the neutron~\cite{Gentile:2011zz}. In recent years there has been fair amount of discussion in the ambiguity of the value of axial dipole mass $M_{A}$ in the dipole parameterization of the axial vector form factor in the case of $\nu_\mu - N$ scattering~\cite{Alvarez-Ruso:2017oui, Katori:2016yel} implying uncertainties in the production cross sections and polarizations of $\tau$ leptons in the case of quasielastic $\nu_{\tau} - N$ scattering which have not been studied.
 
  In this work, we study the effect of using various parametric forms of the weak vector and axial vector form factors as well as the effect of pseudoscalar and second class current form factors in the axial vector sector on the total cross section, differential cross section and the polarization components of $\tau$ leptons produced in the quasielastic $\nu_{\tau}~(\bar{\nu}_{\tau})-$nucleon scattering following our earlier works in the case of $\nu_{\mu}~(\bar{\nu}_{\mu})-$nucleon scattering~\cite{Akbar:2016awk, Fatima:2018tzs, Fatima:2018gjy, Fatima:2018wsy}.
 In section~\ref{mat_element}, we define the matrix element in terms of the various weak form factors and describe their properties under symmetry transformations and express the differential cross sections in terms of these form factors. In section~\ref{polarization}, we discuss the formalism for calculating the polarization observables of the final lepton produced in the quasielastic scattering processes. In section~\ref{total_cross_section},  we present and discuss the numerical results for the total cross section and average polarizations while in section~\ref{cross_section}, the results are presented and discussed for the differential cross sections and $Q^{2}$-dependence of the polarization observables of the $\tau$ leptons and summarize our results with conclusions in section~\ref{summary}.
  
\section{Formalism}\label{mat_element}
 \begin{figure}
 \begin{center}
    \includegraphics[height=4cm,width=16cm]{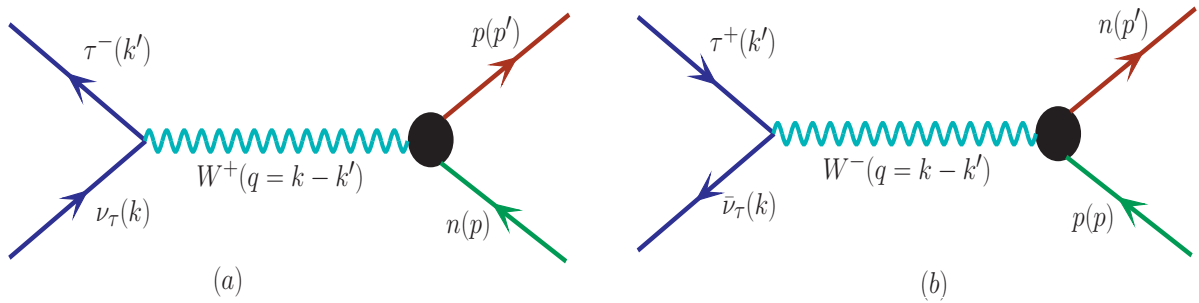}
  \caption{Feynman diagram for the processes (a)~$\nu_{\tau} (k) + n(p) \longrightarrow \tau^{-}(k^{\prime}) + p (p^{\prime})$ and (b)~$\bar{\nu}_\tau (k) + p (p) \rightarrow \tau^+ (k^\prime) + n (p^\prime)$, 
  where the quantities in the bracket represent four momenta of the corresponding particles.}\label{fyn_hyp}
   \end{center}
 \end{figure}
 
\subsection{Matrix element}
The transition matrix element for the processes depicted in Fig.~\ref{fyn_hyp}, given by
\begin{eqnarray}\label{process1}
 \nu_\tau (k) + n (p) &\longrightarrow& \tau^- (k^\prime) + p (p^\prime), ~\\ 
 \label{process2}
\bar{\nu}_\tau (k) + p (p) &\longrightarrow& \tau^+ (k^\prime) + n (p^\prime),    
\end{eqnarray}
 is written as
 \begin{eqnarray}
 \label{matrixelement}
 {\cal{M}} = \frac{G_F}{\sqrt{2}} \cos \theta_{c}~ l^\mu {{J}}_\mu,
 \end{eqnarray}
 where the quantities in the brackets of Eqs.~(\ref{process1}) and (\ref{process2}) represent the four momenta of the respective
 particles, $G_F$ is the Fermi coupling constant and $\theta_c~(=13.1^\circ)$ is the Cabibbo mixing angle. The leptonic 
 current $l^\mu$ is given by
 \begin{equation}\label{l}
 l^\mu = \bar{u} (k^\prime) \gamma^\mu (1 \pm \gamma_5) u (k),
\end{equation}
where $(+)-$ sign is for (anti)neutrino. The hadronic current ${J}_\mu$ is expressed as:
\begin{equation}\label{j}
 {{J}}_\mu =  \bar{u} (p^\prime) {\Gamma_\mu} u (p)
\end{equation}
with
\begin{equation}\label{gamma}
 {\Gamma_\mu} = V_\mu - A_\mu.
\end{equation}
The vector ($V_\mu$) and the axial vector ($A_\mu$) currents are given by~\cite{Fatima:2018tzs, Fatima:2018gjy}:
\begin{eqnarray}\label{vx}
 \langle N^\prime(p^\prime) | V_\mu| N(p) \rangle &=& \bar{u}(p^\prime) \left[ \gamma_\mu f_1(Q^2)+i\sigma_{\mu \nu} 
 \frac{q^\nu}{M_{p}+M_{n}} f_2(Q^2) + \frac{2 ~q_\mu}{M_{p}+M_{n}} f_3(Q^2) \right] u(p),
  \end{eqnarray}
and 
\begin{eqnarray}\label{vy}
  \langle N^\prime(p^\prime) | A_\mu| N(p) \rangle &=& \bar{u} (p^\prime) \left[ \gamma_\mu \gamma_5 g_1(Q^2) + 
  i\sigma_{\mu \nu} \frac{q^\nu}{M_{p}+M_{n}} \gamma_5 g_2(Q^2) + \frac{2 ~q_\mu} {M_{p}+M_{n}} g_3(Q^2) 
  \gamma_5 \right] u(p), 
\end{eqnarray}
where $N,~N^{\prime}$ represents a nucleon $n,p$, with $M_{p}$ and $M_{n}$ being the masses of the proton and the neutron. However, in the numerical calculations we have used $M=\frac{M_{p} + M_{n}}{2}$ in place of $M_{p}$ and $M_{n}$. $q_\mu (= k_\mu - k_\mu^\prime = 
p_\mu^\prime -p_\mu)$ is the four momentum transfer with $Q^2 = - q^2, Q^2 \ge 0$. $f_1 (Q^2)$, $f_2 (Q^2)$ and 
$f_3 (Q^2)$ are the vector, weak magnetic and induced scalar form factors and $g_1 (Q^2)$, $g_2 (Q^2)$ and 
$g_3 (Q^2)$ are the axial vector, induced tensor~(or weak electric) and induced pseudoscalar form factors, 
respectively. According to the classification of G-invariance introduced by Weinberg~\cite{Weinberg:1958ut}, the hadronic current associated with the form factors $f_{1,2} (Q^2)$ and $g_{1,3} (Q^2)$  correspond to the first class currents, while the hadronic current associated with the form factors $f_{3} (Q^2)$ and $g_{2} (Q^2)$ correspond to the second class currents~(SCC) .

\subsection{Weak transition form factors}\label{W_FF}
The general properties of the weak vector and axial vector form factors consistent with the constraints due to the symmetry properties of the 
weak hadronic currents are summarized below~\cite{Marshak, Pais:1971er, LlewellynSmith:1971uhs}:
\begin{itemize}
 \item [(a)] T invariance implies that all the vector and axial vector form factors $f_{1-3} (Q^2)$ and $g_{1-3} (Q^2)$ are real.
 
 \item [(b)] The hypothesis that the weak vector currents and its conjugate along with the isovector part of the 
 electromagnetic current form an isotriplet implies that the weak vector form factors $f_1 (Q^2)$ and $f_2 (Q^2)$ are 
 related to the isovector electromagnetic form factors of the nucleon i.e. $f_{1,2} (Q^2) = f_{1,2}^p (Q^2) - 
 f_{1,2}^n (Q^2)$. The hypothesis ensures conservation of vector current~(CVC) in the weak sector.
 
 \item [(c)] The hypothesis of CVC of the weak vector currents implies that $f_3 (Q^2) = 0$.
 
 \item [(d)] The principle of G-invariance implies the second class current form factors to be zero, {\it i.e.}, $f_3 (Q^2)=0$ and $g_2 (Q^2) =0$.
 
 \item [(e)] The hypothesis of PCAC relates the form factor $g_3 (Q^2)$ to the form factor $g_1 (Q^2)$ through the 
 Goldberger-Treiman~(GT) relation. 

\end{itemize}
 
 In summary, the form factor $f_3 (Q^2) = 0$ both by CVC hypothesis and G-invariance. The form factor $g_2 (Q^2) = 0$ only in the 
 presence of G-invariance. In the case of G-noninvariance, $g_2 (Q^2)$ is non-vanishing and if it is real, it preserves 
 T invariance whereas if it is purely imaginary or complex, the second class currents~(SCC) in the axial vector sector violate T invariance.
 
The expressions for the vector form factors $f_{1,2} (Q^2)$ in 
terms of the electromagnetic form factors $f_{1,2}^p (Q^2)$ and $f_{1,2}^n (Q^2)$ are given as 
\begin{eqnarray}
\label{f12pn}
 f_{1,2}(Q^2)&=&f^p_{1,2}(Q^2)-f^n_{1,2}(Q^2).
 \end{eqnarray}
The electromagnetic form factors $f_{1,2}^p (Q^2)$ and $f_{1,2}^n (Q^2)$ are expressed in terms of the Sachs electric~($G_E^{p,n} (Q^2)$) 
and magnetic~($G_M^{p,n} (Q^2)$) form factors of the nucleons as
\begin{eqnarray}\label{f1pn}
f_1^{p,n}(Q^2)&=&\left(1+\frac{Q^2}{4M^2}\right)^{-1}~\left[G_E^{p,n}(Q^2)+\frac{Q^2}{4M^2}~G_M^{p,n}(Q^2)\right],\\
\label{f2pn}
f_2^{p,n}(Q^2)&=&\left(1+\frac{Q^2}{4M^2}\right)^{-1}~\left[G_M^{p,n}(Q^2)-G_E^{p,n}(Q^2)\right].
\end{eqnarray}
For $G_E^{p,n}(Q^2)$ and $G_M^{p,n}(Q^2)$ various parameterizations are available in the literature~\cite{Bradford:2006yz, Bosted:1994tm, Budd:2004bp,Alberico:2008sz, Kelly:2004hm, Galster:1971kv, Platchkov:1989ch, Punjabi:2015bba} and are given explicitly in Appendix-I. In our 
numerical calculations, we have used the parameterization given by Bradford {\it et al.}~(BBBA05)~\cite{Bradford:2006yz} unless stated otherwise. These nucleon Sachs' form factors have been determined by fitting the data available from the electron-proton scattering experiments on the total cross section and polarization observables.
In modern times, with the advancement in the detector technology, it becomes possible to perform such experiments with high intensity electron beams available at JLab~\cite{Punjabi:2015bba} and MAMI~\cite{MAMI}. Moreover, with such precise experiments Sachs' electric and magnetic form factors have also been determined from the data available from the single and double polarization
measurements. Recently from the JLab
experiment~\cite{Punjabi:2015bba}, the ratio of electric to magnetic form factor of the proton $G_{E}^{p} (Q^2)/G_{M}^{p} (Q^2)$ from the double polarization measurement at $Q^{2} \le$ 5.6 GeV$^2$ has been measured and it shows large deviation from the earlier results. Furthermore, in the case of neutron electric form factor $G_{E}^{n} (Q^2)$, the recent data on the single and double polarization observables are available which are fitted using the different parameterizations of $G_{E}^{n} (Q^{2})$~\cite{Galster:1971kv, Platchkov:1989ch, Kelly:2004hm, Punjabi:2015bba}. We have used these parameterizations of $G_{E}^{n} (Q^2)$ explicitly in our numerical calculations to show the dependence of the total and differential cross sections as well as the polarization observables 
 on $G_{E}^{n} (Q^2)$.

$g_1 (Q^2)$ is the axial vector form factor for $n \rightarrow p$ 
  transition which is determined experimentally from the quasielastic (anti)neutrino scattering from the nucleons and 
  is parameterized as
  \begin{equation}\label{ga}
  g_1 (Q^2) = \frac{g_A (0)}{\left( 1 + \frac{Q^2}{M_A^2} \right)^2},
  \end{equation}
  with $g_A(0) = 1.267$~\cite{Cabibbo:2003cu} and $M_A = 1.026$ GeV~\cite{Bernard:2001rs}. 
  
  The weak electric form factor $g_{2} (Q^2)$ associated with the second class current is 
  parameterized as 
  \begin{equation}\label{g2}
  g_2 (Q^2) = \frac{g_2 (0)}{\left( 1 + \frac{Q^2}{M_2^2} \right)^2},
  \end{equation} 
  in analogy with $g_1 (Q^2)$. In the present work, we have studied the effect of the second class currents assuming
   T invariance and have done numerical calculations by taking $g_{2} (0) = g_{2}^{R} (0)= 0$ and $\pm 1$~\cite{Fatima:2018tzs},
   $M_2= M_A= 1.026$ GeV.
  
  The contribution of $g_3 (Q^2)$ to the matrix element in $\nu_{\tau}~(\bar{\nu}_{\tau}) - N$ scattering 
   is proportional to $m_\tau^{2}$, and becomes significant due to the high value of $m_\tau$. 
  The most widely used parameteric form of the pseudoscalar form factor is obtained using the hypothesis of PCAC along with the GT 
  relation~\cite{Goldberger:1958vp} in the pion pole dominance approximation and is given as~\cite{Fatima:2018tzs}:
  \begin{equation}\label{g3}
  g_3(Q^2)=\frac{2M^2 g_1(Q^2)}{m_{\pi}^2+Q^2},
  \end{equation}
  where $m_{\pi}$ is the pion mass.
  
  Apart from the above relation, there exist other parameterizations for the pseudoscalar form factor in the literature~\cite{Schindler:2006it, Schroder:2001rc}. One of the recent parameterizations of $g_{3}(Q^2)$ is based on PCAC and modified GT relation in which the pseudoscalar form factor is expressed in terms of $g_{1} (Q^2)$ and the pion-nucleon form factor $G_{\pi N} (Q^2)(= \frac{Mg_{A}(0)}{f_{\pi}} - g_{\pi N} \Delta \frac{Q^{2}}{m_{\pi}^{2}})$ and is given by~\cite{Schindler:2006it}
\begin{eqnarray}\label{fp1}
 {{ g_3 (Q^2)}}&=&{{\frac{M}{Q^2}\left[
 \left(\frac{2 m_\pi^2 f_\pi}{m_\pi^2+Q^2}\right)~\left(- \frac{M g_A(0)}{f_\pi}+\frac{g_{\pi N}\Delta Q^2}{m_\pi^2} \right) + 2 M g_1(Q^2) \right].}}
\end{eqnarray}
  where 
   $f_{\pi} = 92.47$~MeV is the pion decay constant and $
\Delta=1-\frac{M g_A (0)}{f_\pi g_{\pi N}}$, with $g_{\pi N}=13.21^{+0.11}_{-0.05}$ \cite{Schroder:2001rc}. 

\subsection{Cross section}
The general expression of the differential cross section for the processes~(\ref{process1}) and (\ref{process2}), in the 
laboratory frame, is given by
 \begin{eqnarray}
 \label{crosv.eq}
 d\sigma&=&\frac{1}{(2\pi)^2}\frac{1}{4 M E_{\nu}}\delta^4(k+p-k^\prime-p^\prime) 
 \frac{d^3k^\prime}{2E_{k^\prime}} \frac{d^3p^\prime}{2E_{p^\prime}} \overline{\sum} \sum |{\cal{M}}|^2,
 \end{eqnarray}
where $E_\nu$ = ($E_{{\bar{\nu}_{\tau}}}$)$E_{\nu_\tau}$ is the incoming (anti)neutrino energy. The transition matrix 
element squared is defined as:
\begin{equation}\label{matrix}
  \overline{\sum} \sum |{\cal{M}}|^2 = \frac{G_F^2 \cos^2 \theta_c}{2} \cal{J}^{\mu \nu} \cal{L}_{\mu \nu},
\end{equation} 

where the hadronic and the leptonic tensors are obtained using Eqs.~(\ref{l}) and (\ref{j}) as
\begin{eqnarray}\label{J}
\cal{J}_{\mu \nu} &=& \overline{\sum} \sum J_\mu J_\nu^\dagger = \frac{1}{2} \mathrm{Tr}\left[\Lambda({p^\prime}) 
\Gamma_{\mu} \Lambda({p}) \tilde{\Gamma}_{\nu} \right], \\
  \label{L}
\cal{L}^{\mu \nu} &=& \overline{\sum} \sum l_\mu l_\nu^\dagger = \mathrm{Tr}\left[\gamma^{\mu}(1 \pm \gamma_{5}) 
\Lambda({k^\prime})\gamma^{\nu}(1 \pm \gamma_{5}) \Lambda({k})~\right],
\end{eqnarray} 
with $\tilde{\Gamma}_{\nu} =\gamma^0 \Gamma^{\dagger}_{\nu} \gamma^0$ and the expression for $\Gamma_\nu$ is given in 
Eq.~(\ref{gamma}). The spin $\frac{1}{2}$ projection operator $\Lambda (P)$ for momentum $P=k, k^\prime,p,p^\prime$ 
corresponding to the initial and the final nucleons and the leptons are given by
\begin{equation}\label{lam}
 \Lambda(P)=(P\!\!\!\!/+M_P),
\end{equation}
where $M_P$ is the mass of the particle with momentum $P$.
 
 Following the above definitions, the differential scattering cross section $d\sigma/dQ^2$ for the processes given in 
 Eqs.~(\ref{process1}) and (\ref{process2}) is written as
\begin{equation}\label{dsig}
 \frac{d\sigma}{dQ^2}=\frac{G_F^2 cos^2\theta_c}{8 \pi {M}^2 {E^2_\nu}} N(Q^2),
\end{equation}
where $N(Q^2) = \cal{J}^{\mu \nu} \cal{L}_{\mu \nu}$ is obtained from the expression given
 in Appendix-A of Ref.~\cite{Fatima:2018tzs} with the substitution of $M^{\prime} =M$ and $m_{\mu}=m_{\tau}$.

\section{Polarization observables of the final lepton}\label{polarization}
 \begin{figure}
 \begin{center}  
        \includegraphics[height=7.5cm,width=13cm]{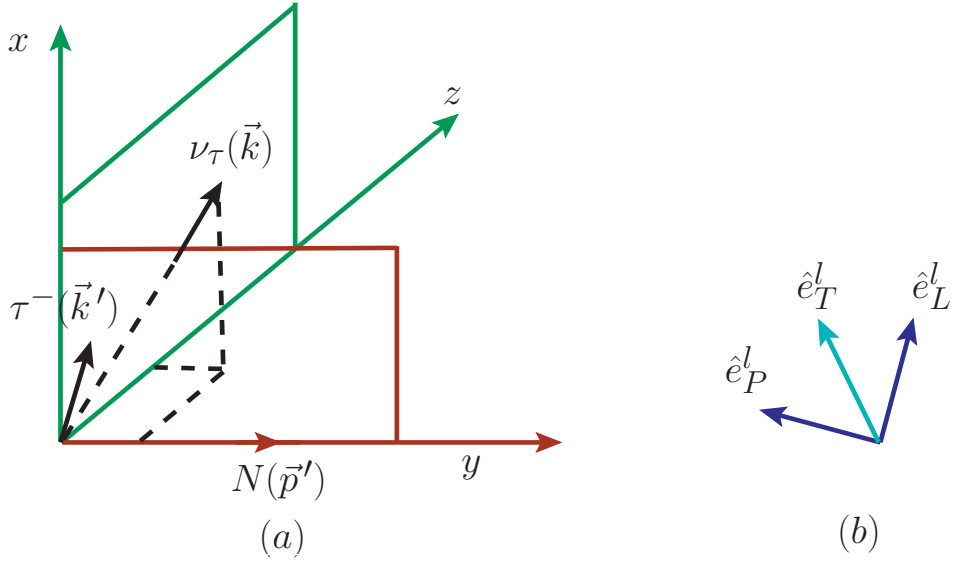}
  \caption{(a) Momentum and polarization directions of the final lepton produced in the reaction ${\nu}_{\tau} (k) + n (p) \longrightarrow \tau^{-} (k^{\prime}) + p (p^{\prime})$. (b)~$\hat{e}_{L}^{l}$, 
  $\hat{e}_{P}^{l}$ and $\hat{e}_{T}^{l}$ represent the orthogonal unit vectors corresponding to the longitudinal, 
  perpendicular and transverse directions with respect to the momentum of the final lepton.}\label{T invariance}
   \end{center}
 \end{figure}

Using the covariant density matrix formalism, the polarization 4-vector~($\zeta^\tau$) of the $\tau$ lepton produced in the 
final state in reactions~(\ref{process1}) and (\ref{process2}) is written as~\cite{Bilenky}
\begin{equation}\label{polarl}
\zeta^{\tau}=\frac{\mathrm{Tr}[\gamma^{\tau}\gamma_{5}~\rho_{f}(k^\prime)]}
{\mathrm{Tr}[\rho_{f}(k^\prime)]},
\end{equation}
and the spin density matrix for the final lepton $\rho_f(k^\prime)$ is given by 
\begin{equation}\label{polar1l}
 \rho_{f}(k^\prime)= {\cal J}^{\alpha \beta}  \text{ Tr}[\Lambda(k') \gamma_\alpha (1 \pm \gamma_5) \Lambda(k) 
 \tilde\gamma_ {\beta} (1 \pm \tilde\gamma_5)\Lambda(k')], 
\end{equation} 
with $\tilde{\gamma}_{\alpha} =\gamma^0 \gamma^{\dagger}_{\alpha} \gamma^0$ and $\tilde{\gamma}_{5} =\gamma^0 
\gamma^{\dagger}_{5} \gamma^0$.

Using the following relations:
\begin{equation}\label{polar3}
\Lambda(k')\gamma^{\tau}\gamma_{5}\Lambda(k')=2m_{\tau} \left(g^{\tau\sigma}-\frac{k'^{\tau}k'^{\sigma}}{m_{\tau}^{2}}
\right)\Lambda(k')\gamma_{\sigma}\gamma_{5}
\end{equation}
and
\begin{equation}\label{polar31}
 \Lambda(k^\prime)\Lambda(k^\prime) = 2m_{\tau} \Lambda(k^\prime),
\end{equation}
$\zeta^\tau$ defined in Eq.~(\ref{polarl}) may also be rewritten as
\begin{equation}\label{polar4l}
\zeta^{\tau}=\left( g^{\tau\sigma}-\frac{k'^{\tau}k'^{\sigma}}{m_\tau^2}\right)
\frac{  {\cal J}^{\alpha \beta}  \mathrm{Tr}
\left[\gamma_{\sigma}\gamma_{5}\Lambda(k') \gamma_\alpha (1 \pm \gamma_5) \Lambda(k) \tilde\gamma_ {\beta} (1 \pm 
\tilde\gamma_5) \right]}
{ {\cal J}^{\alpha \beta} \mathrm{Tr}\left[\Lambda(k') \gamma_\alpha (1 \pm \gamma_5) \Lambda(k) \tilde\gamma_ {\beta} 
(1 \pm \tilde\gamma_5) \right]},
\end{equation}
where $m_\tau$ is the mass of the $\tau$ lepton. In Eq.~(\ref{polar4l}), the denominator is directly related to the differential 
cross section given in Eq.~(\ref{dsig}).

With ${\cal J}^{\alpha \beta}$ and ${\cal L}_{\alpha \beta}$ given in Eqs.~(\ref{J}) and (\ref{L}), respectively, an 
expression for $\zeta^\tau$ is obtained. In the laboratory frame where the initial nucleon is at rest, the polarization 
vector $\vec{\zeta}$, assuming T invariance, is calculated to be a function of 3-momenta of incoming (anti)neutrino $({\vec{k}})$ and outgoing 
lepton $({\vec{k}}\,^{\prime})$, and is given as  
\begin{equation}\label{3poll}
 \vec{\zeta} =\left[{A^l(Q^2)\vec{ k}} + B^l(Q^2){\vec{k}}\,^{\prime} \right], 
\end{equation}
where the expressions of $A^l(Q^2)$ and $B^l(Q^2)$ are obtained from the expression given in Appendix-B of Ref.~\cite{Fatima:2018tzs} with the substitution $M^{\prime} =M$ and $m_{\mu}=m_{\tau}$.

One may expand the polarization vector $\vec{\zeta}$ along the orthogonal directions, ${\hat{e}}_L^l$, ${\hat{e}}_P^l$ 
and ${\hat{e}_T^l}$ in the reaction plane corresponding to the longitudinal, perpendicular and transverse directions of the final lepton~($l$), as depicted in Fig.~\ref{T invariance} and 
defined as
\begin{equation}\label{vectorsl}
\hat{e}_{L}^l=\frac{\vec{ k}^{\, \prime}}{|\vec{ k}^{\,\prime}|},\qquad
\hat{e}_{P}^l=\hat{e}_{L}^l \times \hat{e}_T^l,\qquad   {\rm where}~~~~~ 
\hat{e}_T^l=\frac{\vec{ k}\times \vec{ k}^{\,\prime}}{|\vec{ k}\times \vec{ k}^{\,\prime}|}.
\end{equation}
We then write $\vec{\zeta}$ as:
 \begin{equation}\label{polarLabl}
\vec{\zeta}=\zeta_{L} \hat{e}_{L}^l+\zeta_{P} \hat{e}_{P}^l + \zeta_{T} \hat{e}_{T}^l,
\end{equation}
such that the longitudinal and  perpendicular components of the $\vec{\zeta}$ in the laboratory frame are 
given by
\begin{equation}\label{PLl}
 \zeta_L(Q^2)=\vec{\zeta} \cdot \hat{e}_L^l,\qquad \zeta_P(Q^2)= \vec{\zeta} \cdot \hat{e}_P^l.
\end{equation}
From Eq.~(\ref{PLl}), the longitudinal $P_L (Q^2)$ and perpendicular $P_P(Q^2)$  
components of the polarization vector defined in the rest frame of the outgoing lepton are given by 
\begin{equation}\label{PL1l}
 P_L(Q^2)=\frac{m_\tau}{E_{k^\prime}} \zeta_L(Q^2), \qquad P_P(Q^2)=\zeta_P(Q^2),
\end{equation}
where $\frac{m_\tau}{E_{k^\prime}}$ is the Lorentz boost factor along ${\vec k}\, ^\prime$. Using Eqs.~(\ref{3poll}), 
(\ref{vectorsl}) and (\ref{PLl}) in Eq. (\ref{PL1l}), the longitudinal $P_L(Q^2)$ and perpendicular 
$P_P (Q^2)$ components are calculated to be
\begin{eqnarray}
  P_L (Q^2) &=& \frac{m_\tau}{E_{k^{\prime}}} \frac{A^l(Q^2) \vec{k}.\vec{k}^{\,\prime} + B^l (Q^2) 
  |\vec{k}^{\,\prime}|^2}{N(Q^2)~|\vec{k}^{\,\prime}|},\label{Pll} \\
 P_P (Q^2) &=& \frac{A^l(Q^2) [|\vec{k}|^2 |\vec{k}^{\,\prime}|^2 - (\vec{k}.\vec{k}^{\,\prime})^2]}{N(Q^2)~
 |\vec{k}^{\,\prime}| ~ |\vec{k}\times \vec{k}^{\,\prime}|}.\label{Ppl}
\end{eqnarray}

\section{Results and discussion: Total scattering cross section and average polarizations}\label{total_cross_section}
To study 
the dependence of the cross section $\sigma(E_{\nu_\tau (\bar{\nu}_\tau)})$ and average polarizations $\overline{P}_{L,P} (E_{\nu_{\tau}(\bar{\nu}_{\tau})})$ on the (anti)neutrino's energy $E_{\nu_\tau(\bar{\nu}_\tau)}$, we have integrated $d\sigma/dQ^2$ and $P_{L,P} (Q^2)$ over $Q^2$ and obtained the expressions for $\sigma 
(E_{\nu_\tau(\bar{\nu}_\tau)})$ and $\overline{P}_{L,P} (E_{\nu_{\tau}(\bar{\nu}_{\tau})})$ as:
\begin{eqnarray}\label{total_sig}
\sigma (E_{\nu_\mu(\bar{\nu}_\mu)}) &=& \int_{Q^2_{min}}^{Q^2_{max}} \frac{d\sigma}{dQ^2} dQ^2,
\end{eqnarray}
and 
\begin{eqnarray}\label{average_Plpt}
 \overline{P}_{L,P} (E_{\nu_\tau(\bar{\nu}_\tau)}) &=& \frac{\int_{Q^2_{min}}^{Q^2_{max}} P_{L,P} (Q^2) 
 \frac{d\sigma}{dQ^2} dQ^2}{\int_{Q^2_{min}}^{Q^2_{max}} \frac{d\sigma}{dQ^2} dQ^2}.
\end{eqnarray}
We have used Eq.~(\ref{dsig}) to numerically evaluate the differential cross section $d \sigma/d Q^2$ and Eqs.~(\ref{Pll}) and (\ref{Ppl}) to evaluate the longitudinal $P_L(Q^2)$ and perpendicular $P_P(Q^2)$  
components of polarization of $\tau$ lepton, respectively. The Dirac and Pauli form factors $f_{1,2}^{N} (Q^2); ~ (N=p,n)$ are 
expressed in terms of the electric and magnetic Sachs' form factors, for which the parameterization given by Bradford 
et al.~\cite{Bradford:2006yz} have been used, unless stated otherwise. For $g_1 (Q^2)$ and 
$g_2 (Q^2)$, dipole parameterizations have been used as written in Eqs.~(\ref{ga}) and (\ref{g2}), respectively, with $M_A = $ 
1.026 GeV and $M_2 = M_A$. For $g_3 (Q^2)$, the expressions given in Eqs.~(\ref{g3}) and (\ref{fp1}) have been used. 
\begin{figure}
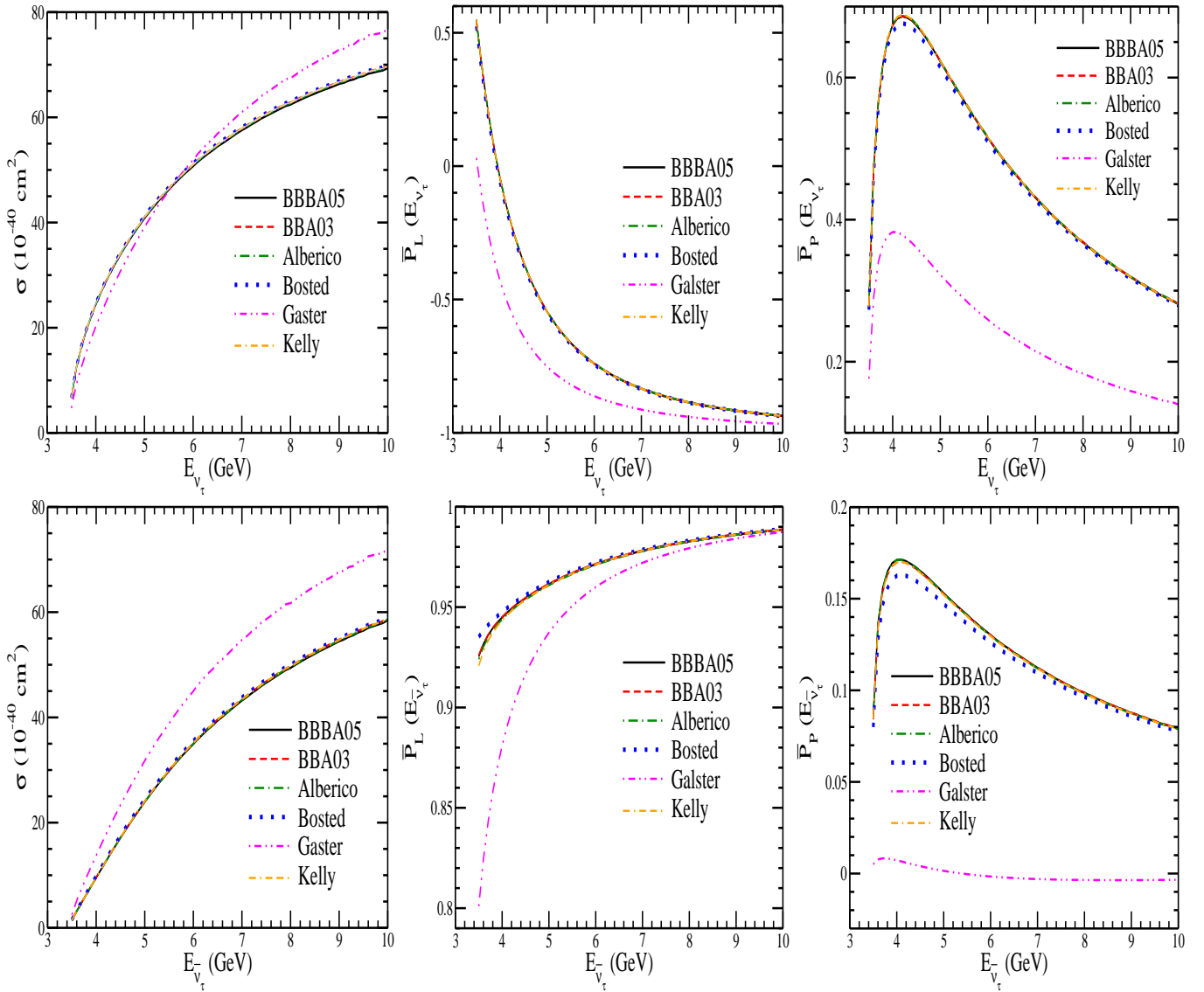

\centering
\includegraphics[height=7.5cm,width=5.9cm]{total_sigma_proton_VFF.eps}
\includegraphics[height=7.5cm,width=5.9cm]{total_Pl_proton_VFF.eps}
\includegraphics[height=7.5cm,width=5.9cm]{total_Pp_proton_VFF.eps}
\includegraphics[height=7.5cm,width=5.9cm]{total_sigma_neutron_VFF.eps}
\includegraphics[height=7.5cm,width=5.9cm]{total_Pl_neutron_VFF.eps}
\includegraphics[height=7.5cm,width=5.9cm]{total_Pp_neutron_VFF.eps}
\caption{(Top panel) Left to Right:  $\sigma$ vs $E_{\nu_{\tau}}$, $\overline{P}_{L} (E_{\nu_{\tau}})$ vs $E_{\nu_{\tau}}$,
 and $\overline{P}_{P} (E_{\nu_{\tau}})$ vs $E_{\nu_{\tau}}$ for the $\nu_{\tau} + n \rightarrow \tau^{-} + p$ process.
 (Bottom panel) Left to Right:  $\sigma$ vs $E_{\bar{\nu}_{\tau}}$, $\overline{P}_{L} (E_{\bar{\nu}_{\tau}})$ vs $E_{\bar{\nu}_{\tau}}$,
 and $\overline{P}_{P} (E_{\bar{\nu}_{\tau}})$ vs $E_{\bar{\nu}_{\tau}}$ for the $\bar{\nu}_{\tau} + p \rightarrow \tau^{+} + n$ process.
   The calculations have been performed using the axial dipole mass $M_{A}=1.026$~GeV and for the different parameterizations of the nucleon vector form factors {\it viz.}, BBBA05~\cite{Bradford:2006yz}~(solid line), BBA03~\cite{Budd:2004bp}~(dashed line), Alberico~\cite{Alberico:2008sz}~(dashed-dotted line), Bosted~\cite{Bosted:1994tm}~(dotted line), 
Galster~\cite{Galster:1971kv}~(double-dotted-dashed line) and Kelly~\cite{Kelly:2004hm}~(double-dashed-dotted line) being used in Eqs.~(\ref{f1pn}) and (\ref{f2pn}).}\label{sigma_VFF_nu}
\end{figure}
\begin{figure}
\centering
\includegraphics[height=7.5cm,width=5.9cm]{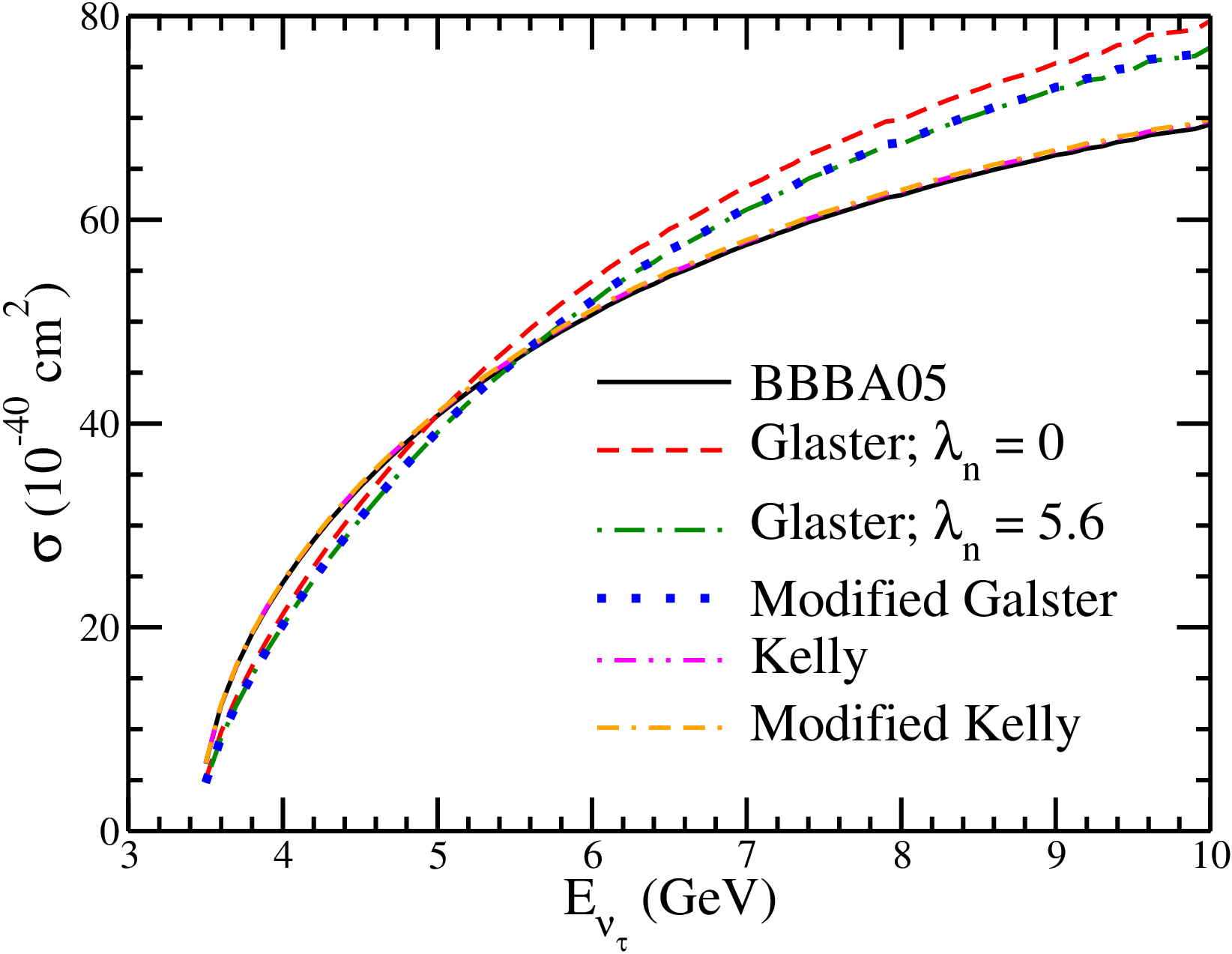}
\includegraphics[height=7.5cm,width=5.9cm]{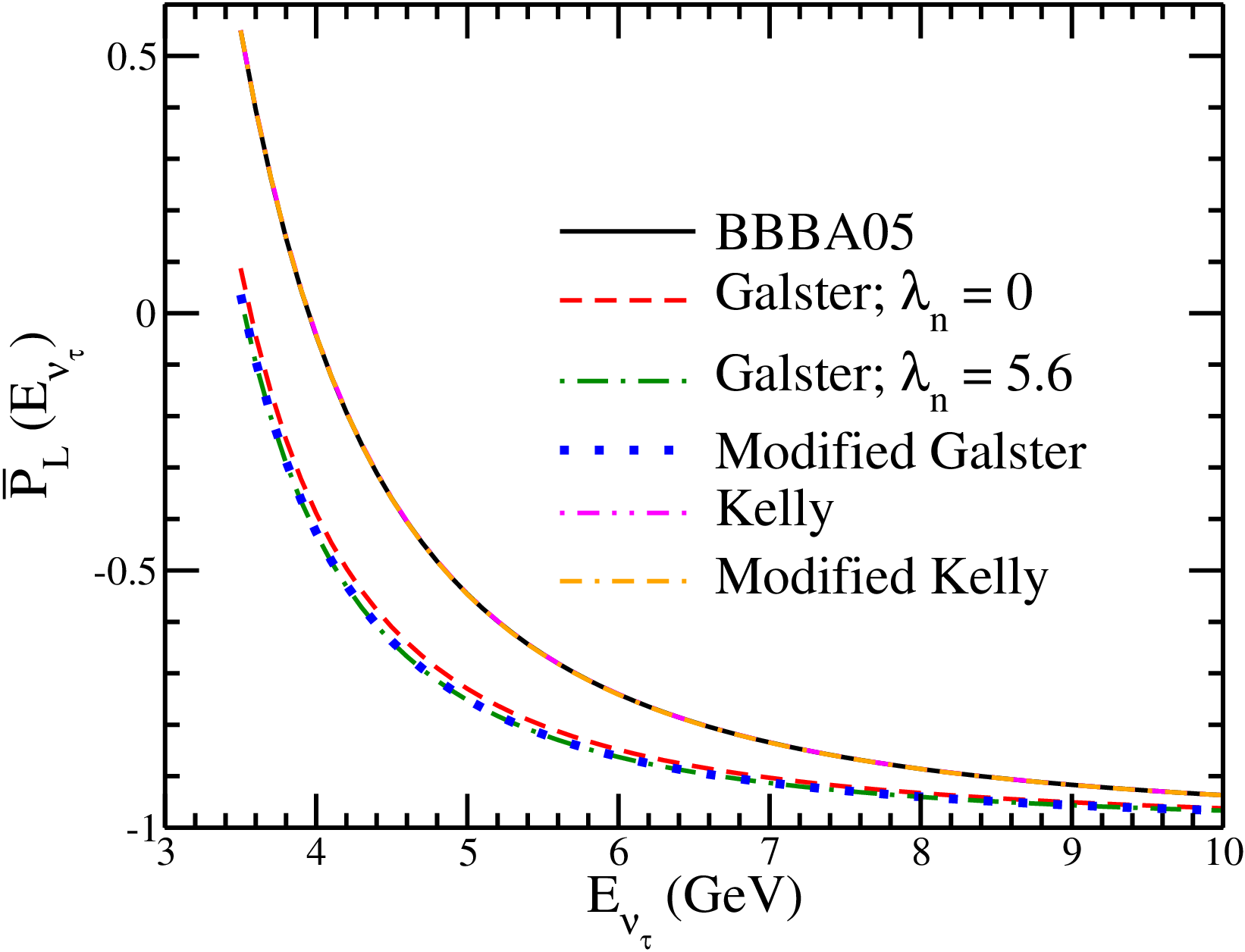}
\includegraphics[height=7.5cm,width=5.9cm]{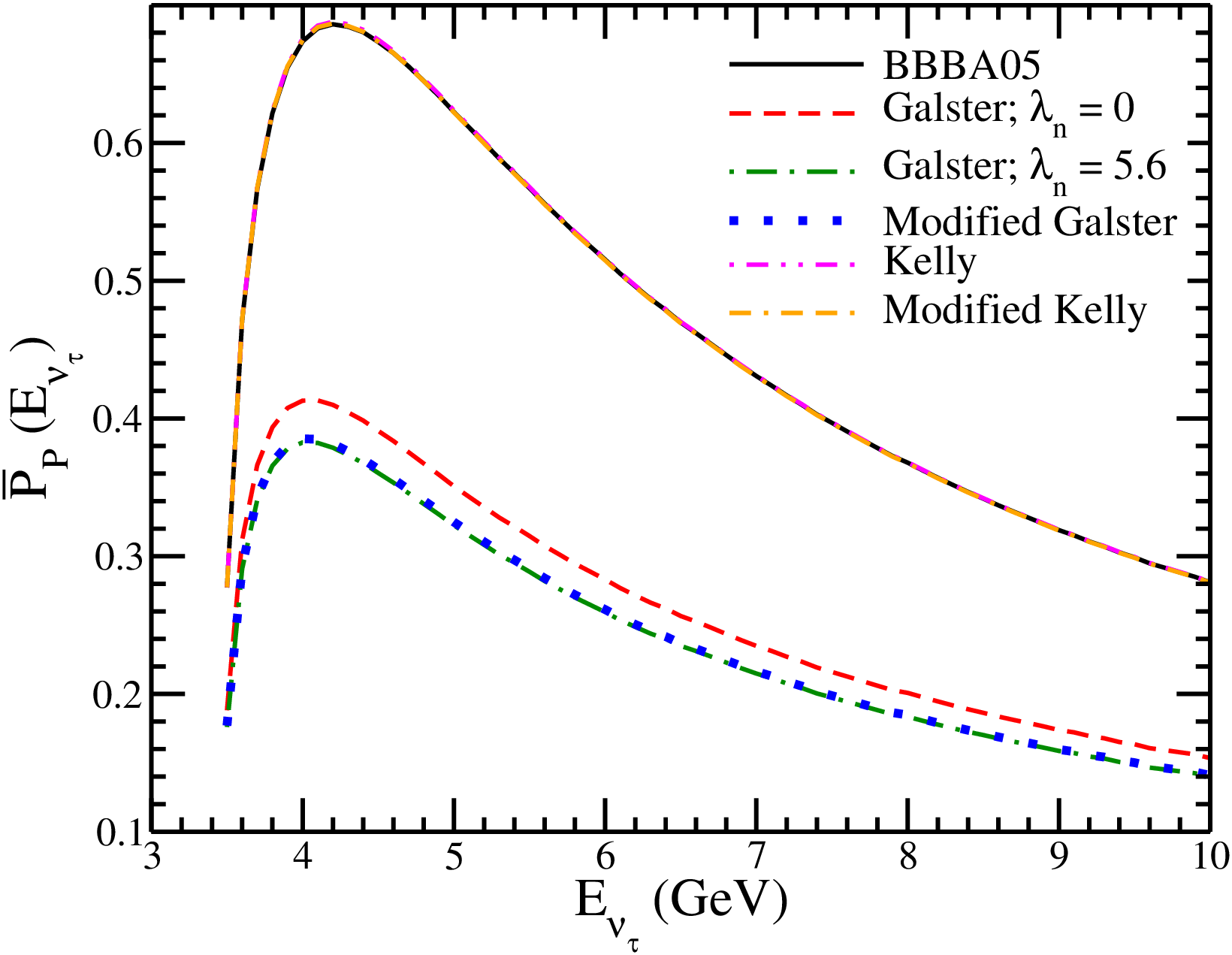}
\includegraphics[height=7.5cm,width=5.9cm]{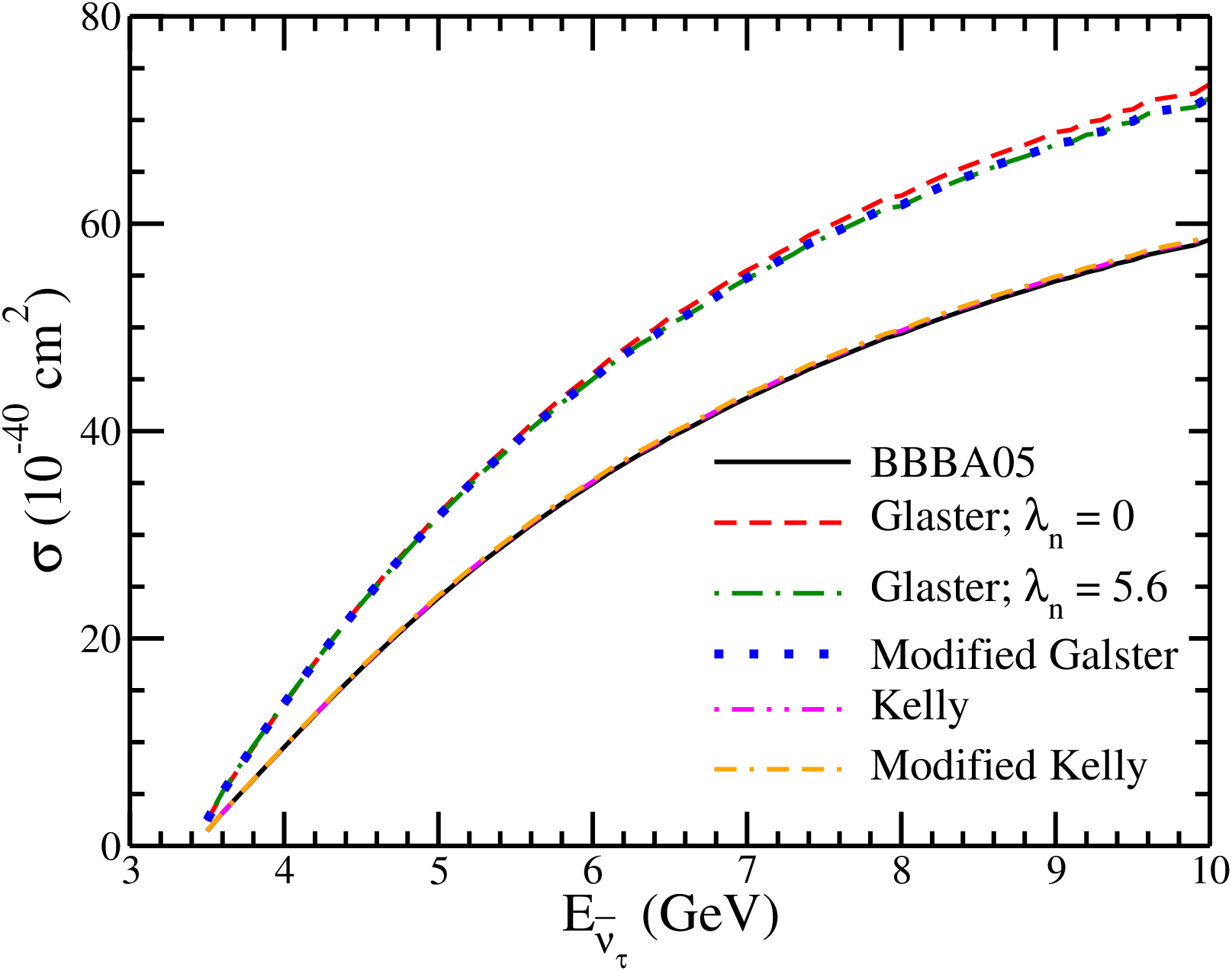}
\includegraphics[height=7.5cm,width=5.9cm]{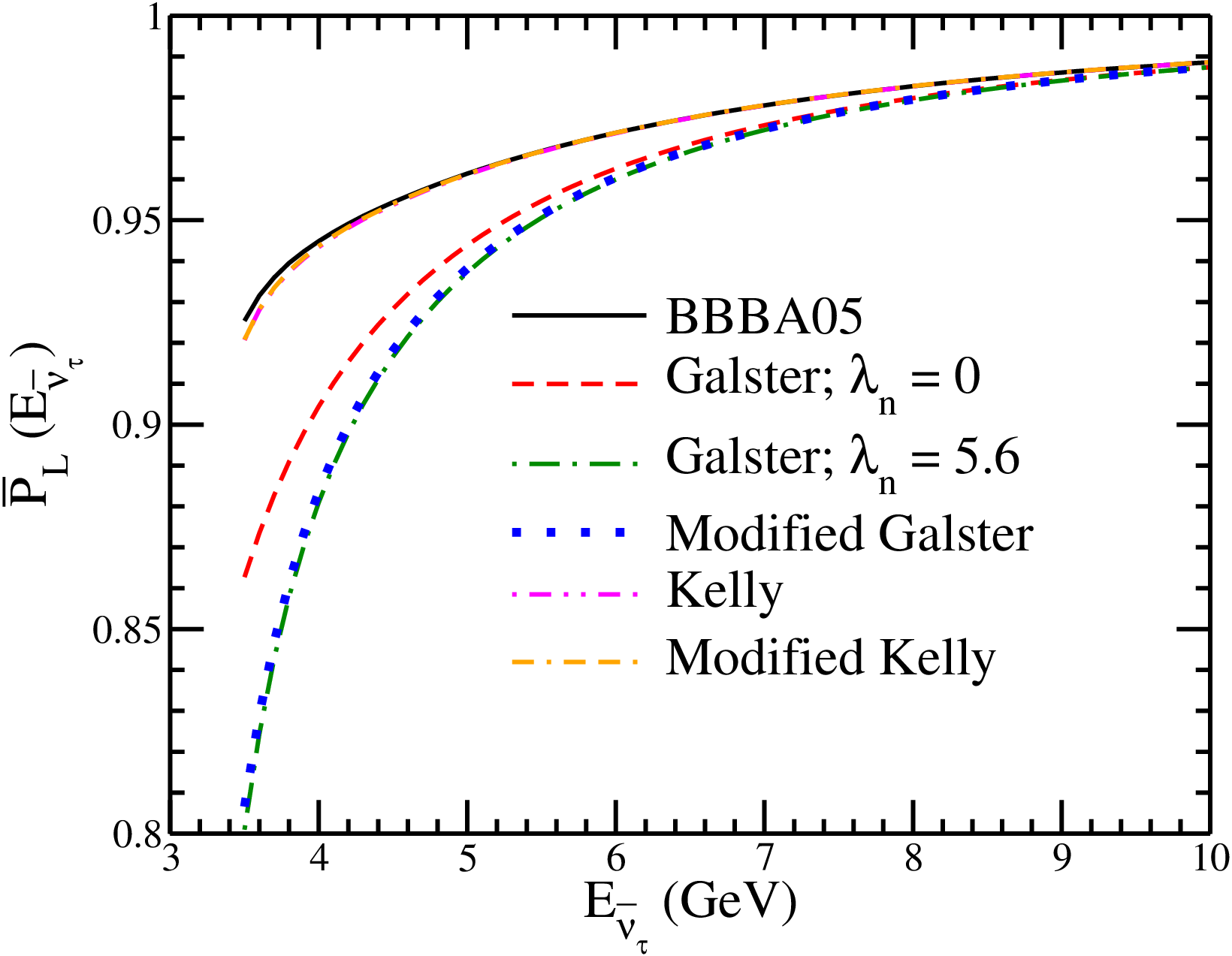}
\includegraphics[height=7.5cm,width=5.9cm]{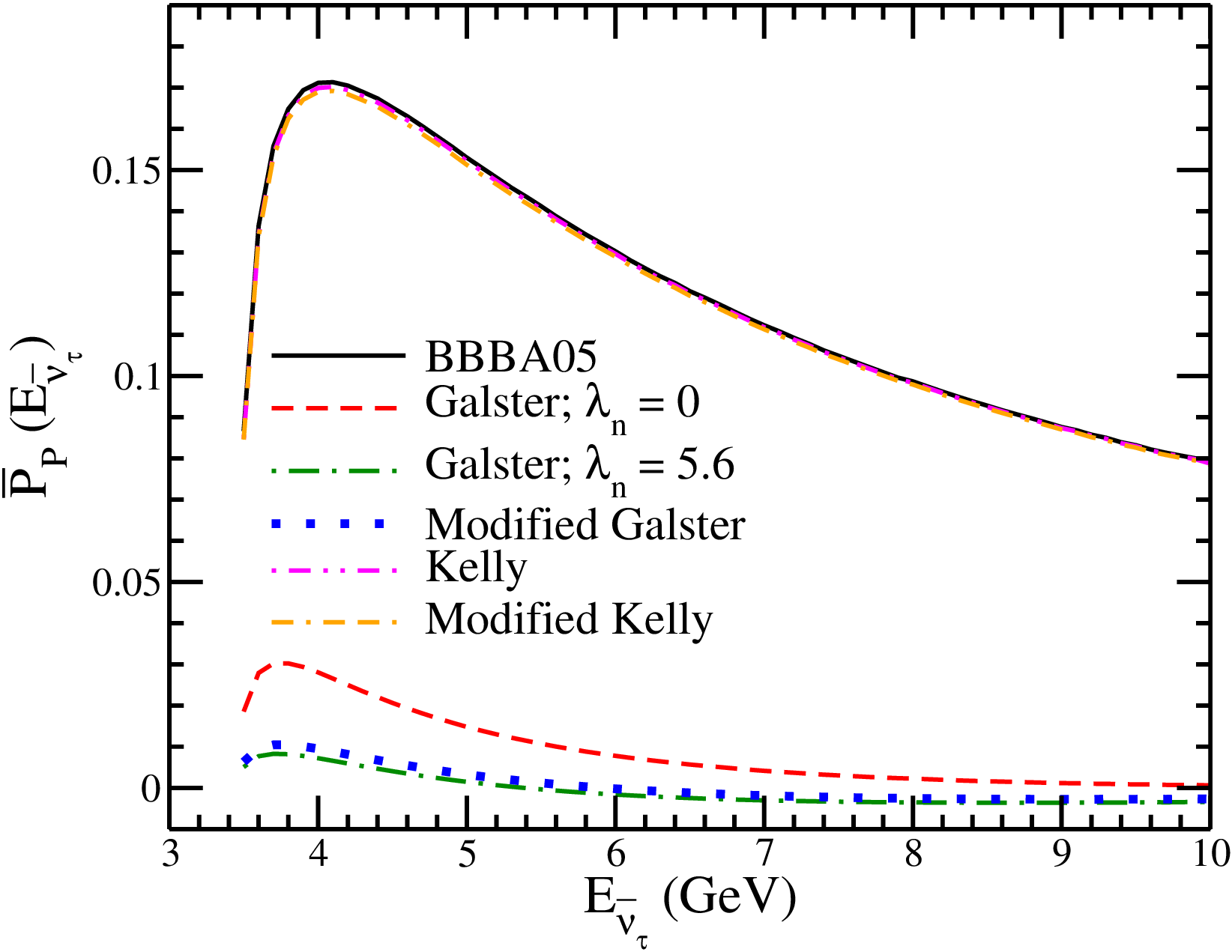}
\caption{(Top panel) Left to Right:  $\sigma$ vs $E_{\nu_{\tau}}$, $\overline{P}_{L} (E_{\nu_{\tau}})$ vs $E_{\nu_{\tau}}$,
 and $\overline{P}_{P} (E_{\nu_{\tau}})$ vs $E_{\nu_{\tau}}$ for the $\nu_{\tau} + n \rightarrow \tau^{-} + p$ process.
 (Bottom panel) Left to Right:  $\sigma$ vs $E_{\bar{\nu}_{\tau}}$, $\overline{P}_{L} (E_{\bar{\nu}_{\tau}})$ vs $E_{\bar{\nu}_{\tau}}$,
 and $\overline{P}_{P} (E_{\bar{\nu}_{\tau}})$ vs $E_{\bar{\nu}_{\tau}}$ for the $\bar{\nu}_{\tau} + p \rightarrow \tau^{+} + n$ process.
   The calculations have been performed using the axial dipole mass $M_{A}=1.026$~GeV and with the different parameterizations of the  neutron Sachs electric form factor $G_E^n (Q^2)$ {\it viz.} BBBA05~\cite{Bradford:2006yz}~(solid line), Galster with $\lambda_n$ = 0~\cite{Galster:1971kv}~(dashed line), Galster with $\lambda_n$ = 5.6~\cite{Galster:1971kv}~(dashed-dotted line), Galster modified given by Platchkov {\it et al.}~\cite{Platchkov:1989ch}~(dotted line), Kelly~\cite{Kelly:2004hm}~(double dotted-dashed line) and Kelly modified given by Punjabi {\it et al.}~\cite{Punjabi:2015bba}~(double dashed-dotted line).}\label{sigma_Gen_nu}
\end{figure}

\begin{figure}
\centering
\includegraphics[height=7.5cm,width=5.9cm]{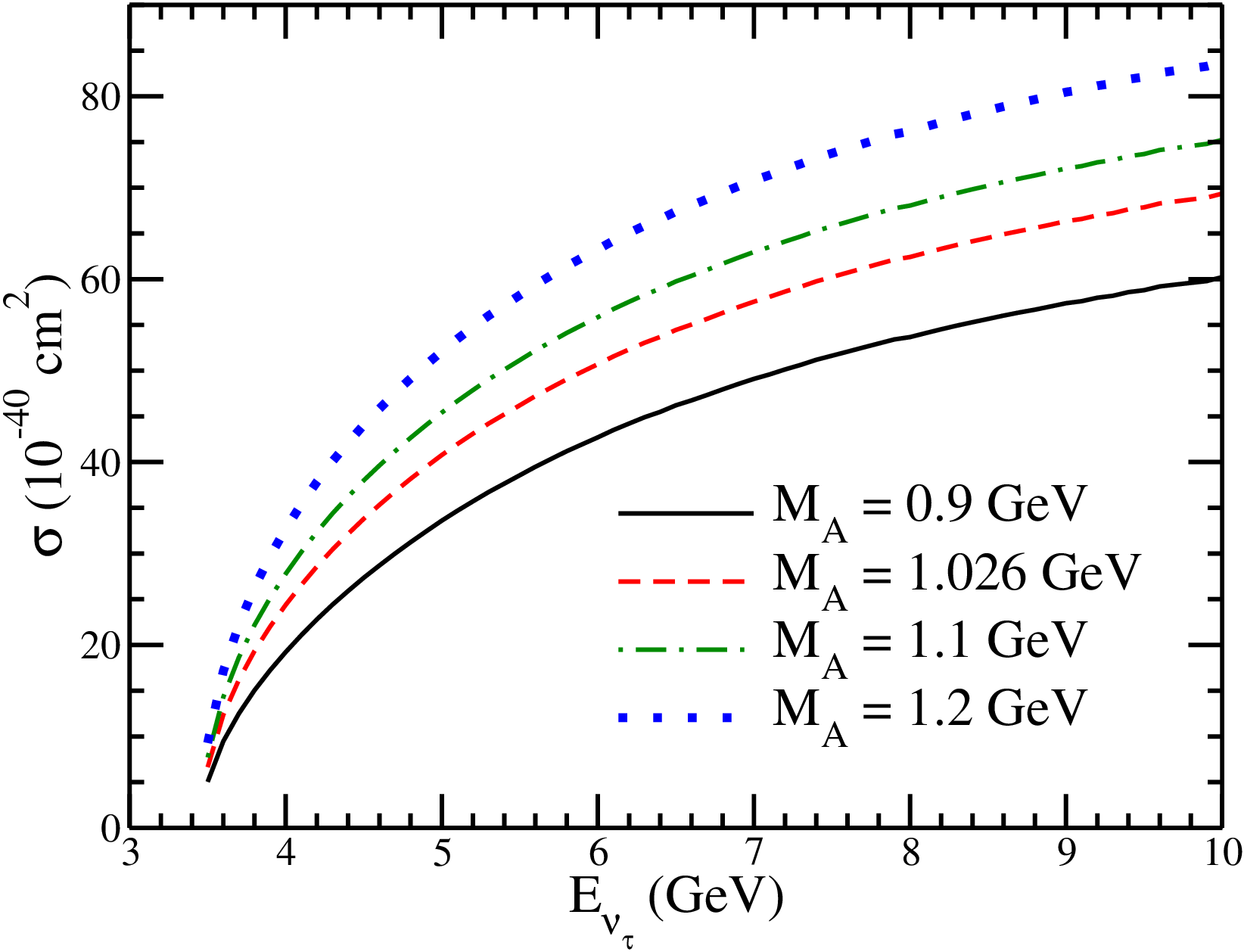}
\includegraphics[height=7.5cm,width=5.9cm]{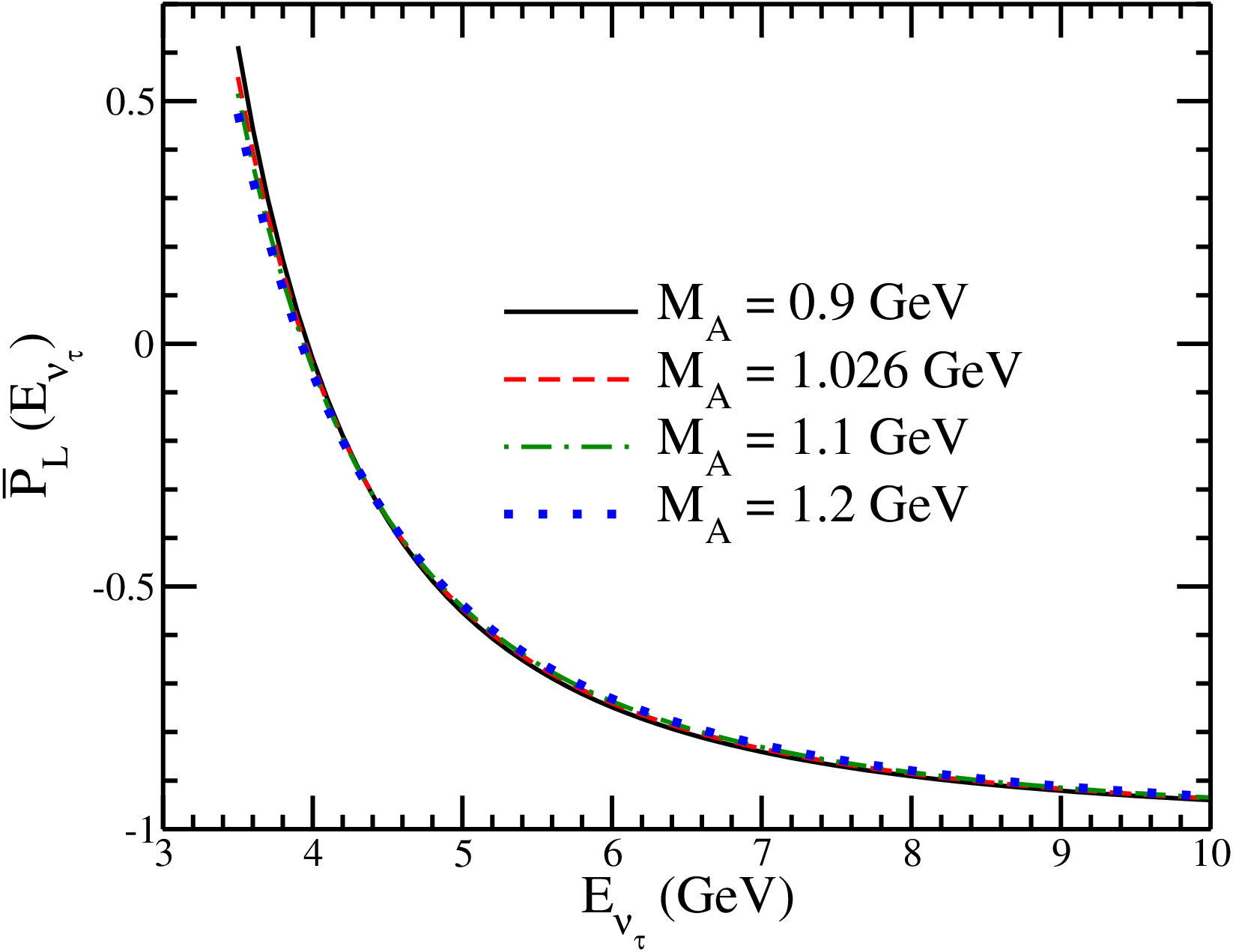}
\includegraphics[height=7.5cm,width=5.9cm]{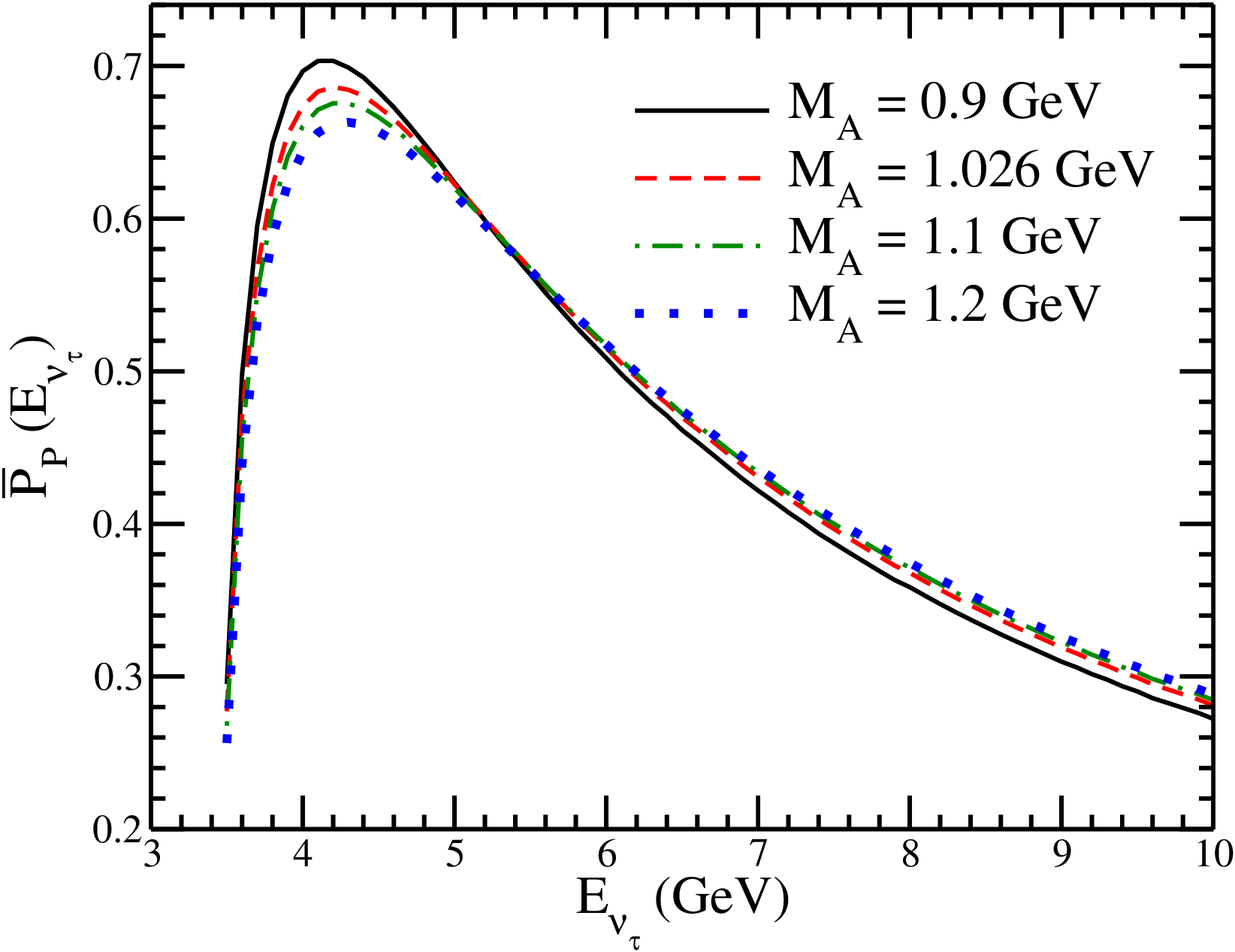}
\includegraphics[height=7.5cm,width=5.9cm]{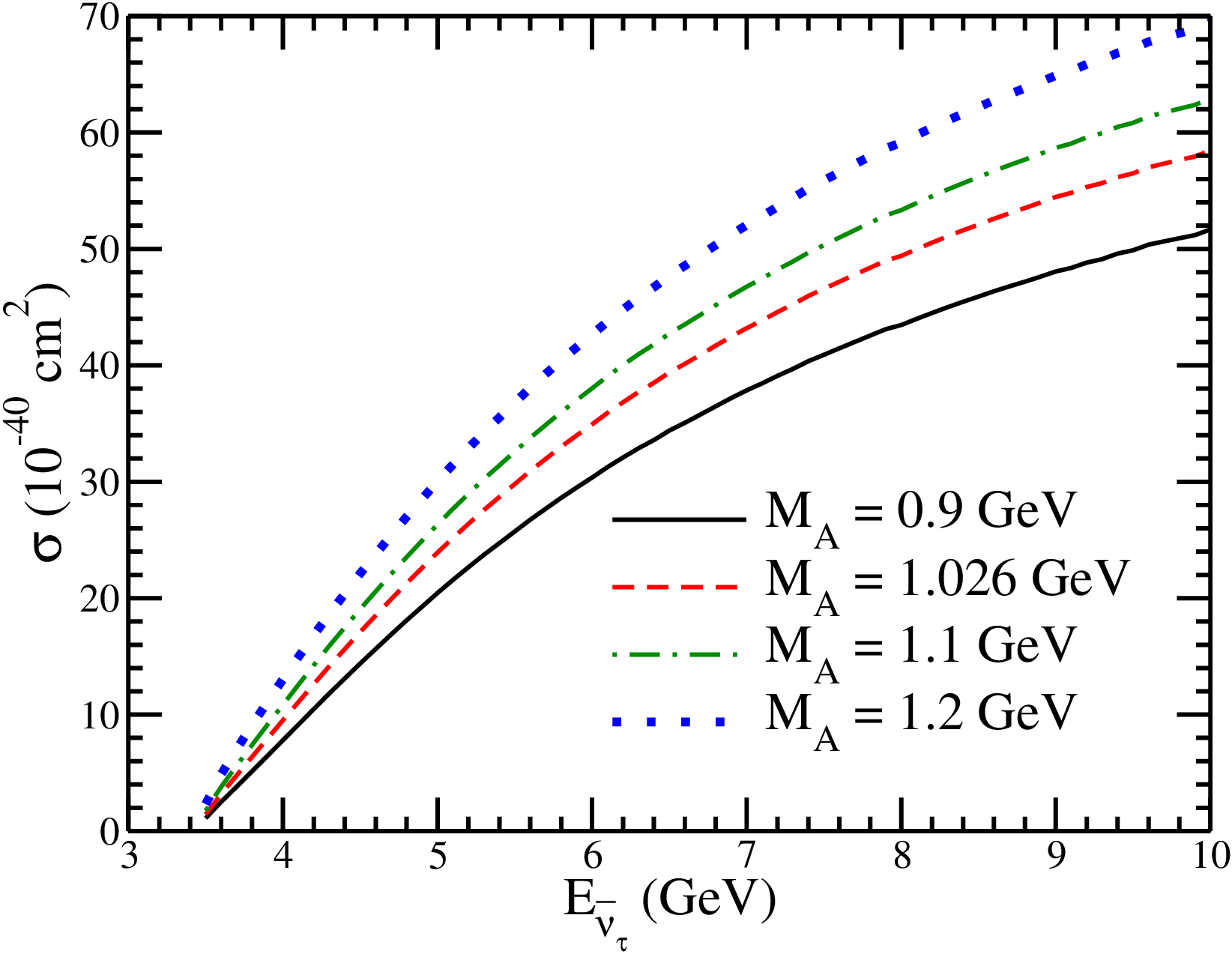}
\includegraphics[height=7.5cm,width=5.9cm]{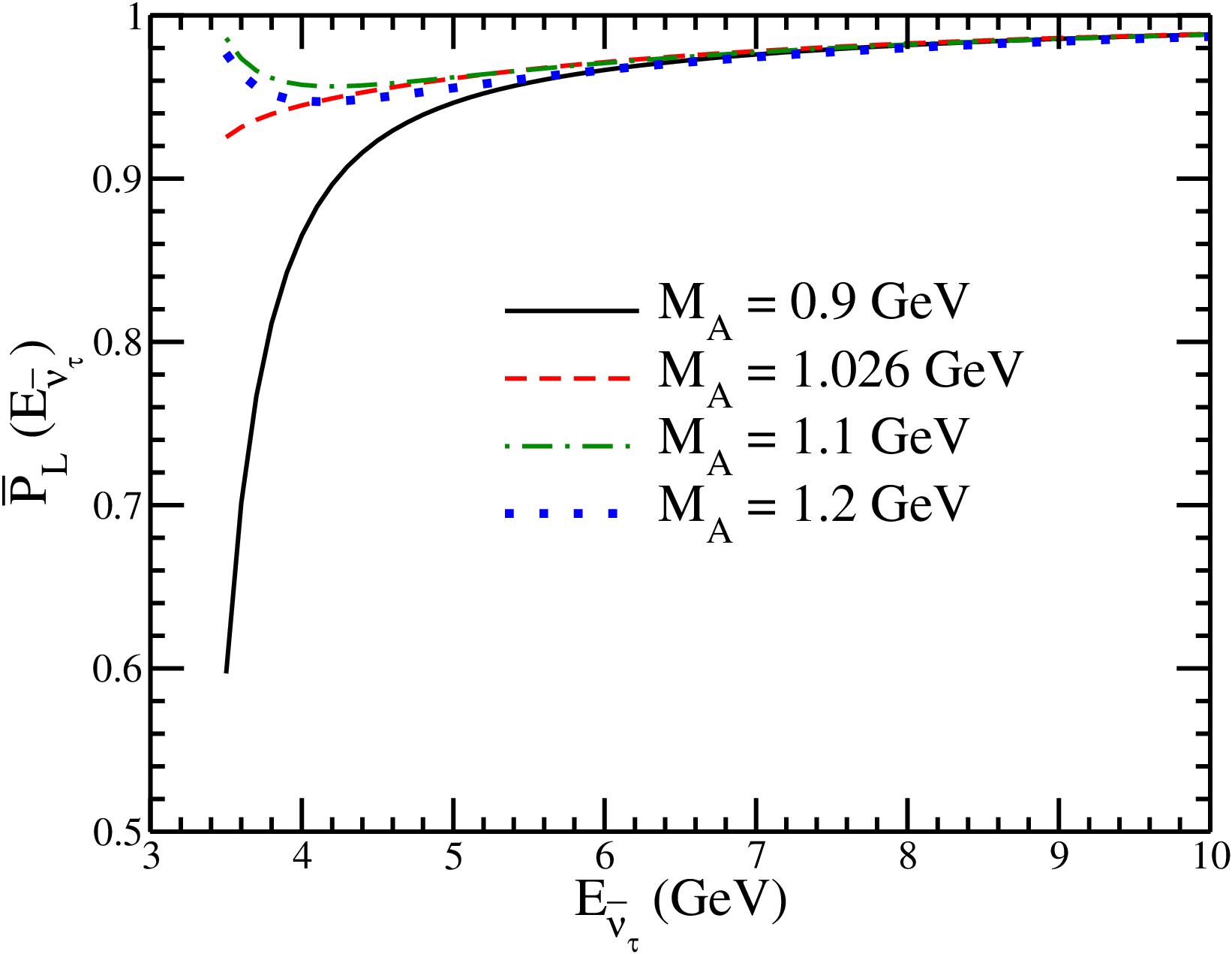}
\includegraphics[height=7.5cm,width=5.9cm]{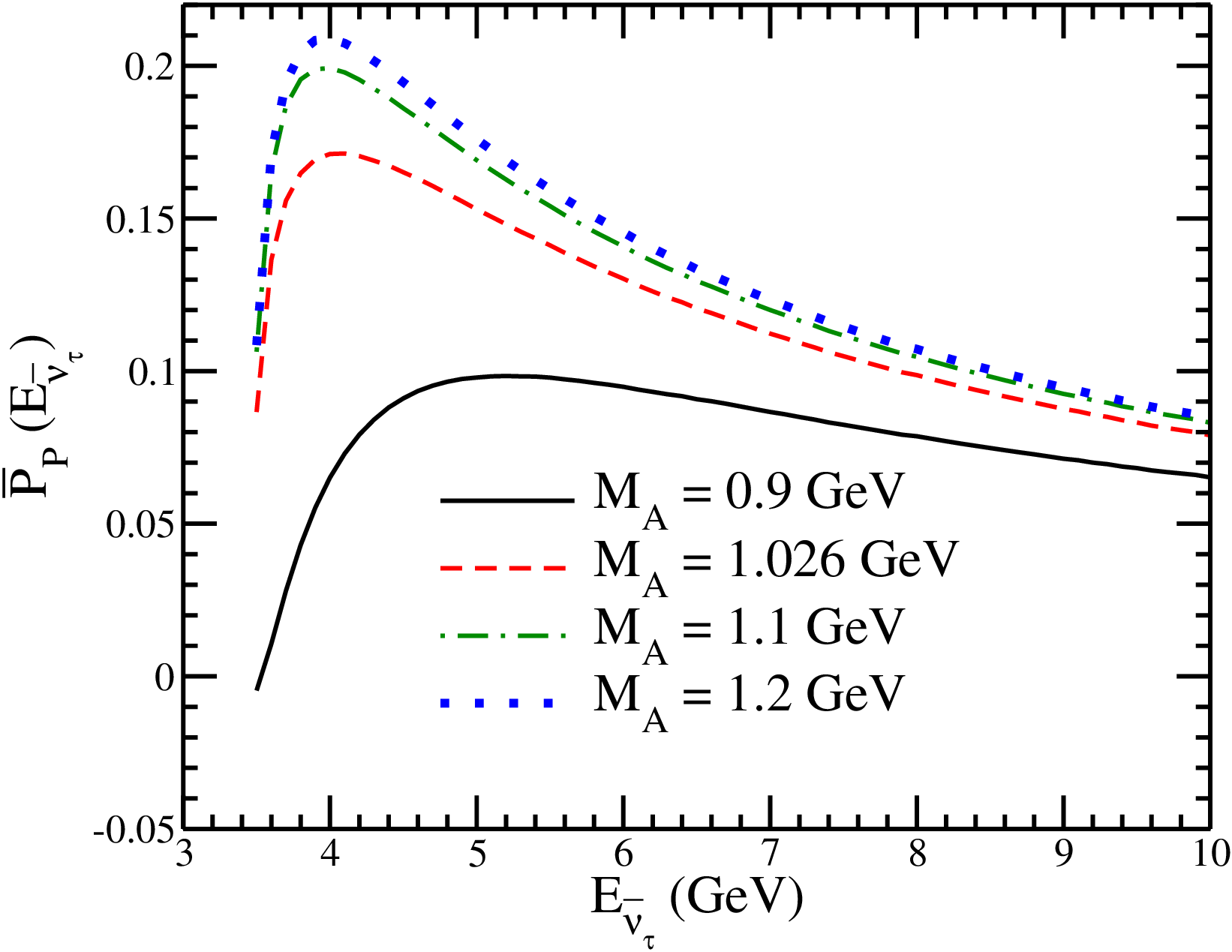}
\caption{(Top panel) Left to Right:  $\sigma$ vs $E_{\nu_{\tau}}$, $\overline{P}_{L} (E_{\nu_{\tau}})$ vs $E_{\nu_{\tau}}$,
 and $\overline{P}_{P} (E_{\nu_{\tau}})$ vs $E_{\nu_{\tau}}$ for the $\nu_{\tau} + n \rightarrow \tau^{-} + p$ process.
 (Bottom panel) Left to Right:  $\sigma$ vs $E_{\bar{\nu}_{\tau}}$, $\overline{P}_{L} (E_{\bar{\nu}_{\tau}})$ vs $E_{\bar{\nu}_{\tau}}$,
 and $\overline{P}_{P} (E_{\bar{\nu}_{\tau}})$ vs $E_{\bar{\nu}_{\tau}}$ for the $\bar{\nu}_{\tau} + p \rightarrow \tau^{+} + n$ process. The calculations have been performed using electric and magnetic Sachs form factors(Eqs.~(\ref{f1pn}) and (\ref{f2pn})) parameterized by Bradford {\it et al.}~\cite{Bradford:2006yz} and for the axial form factor (Eq.~(\ref{ga})), the different dipole mass has been used {\it viz.} $M_{A} =$ 0.9 GeV~(solid line), 1.026 GeV~(dashed line), 1.1 GeV~(dashed-dotted line) and 1.2 GeV~(dotted line).}\label{sigma_MA_nu}
\end{figure}

\begin{figure}
\centering
\includegraphics[height=7.5cm,width=5.9cm]{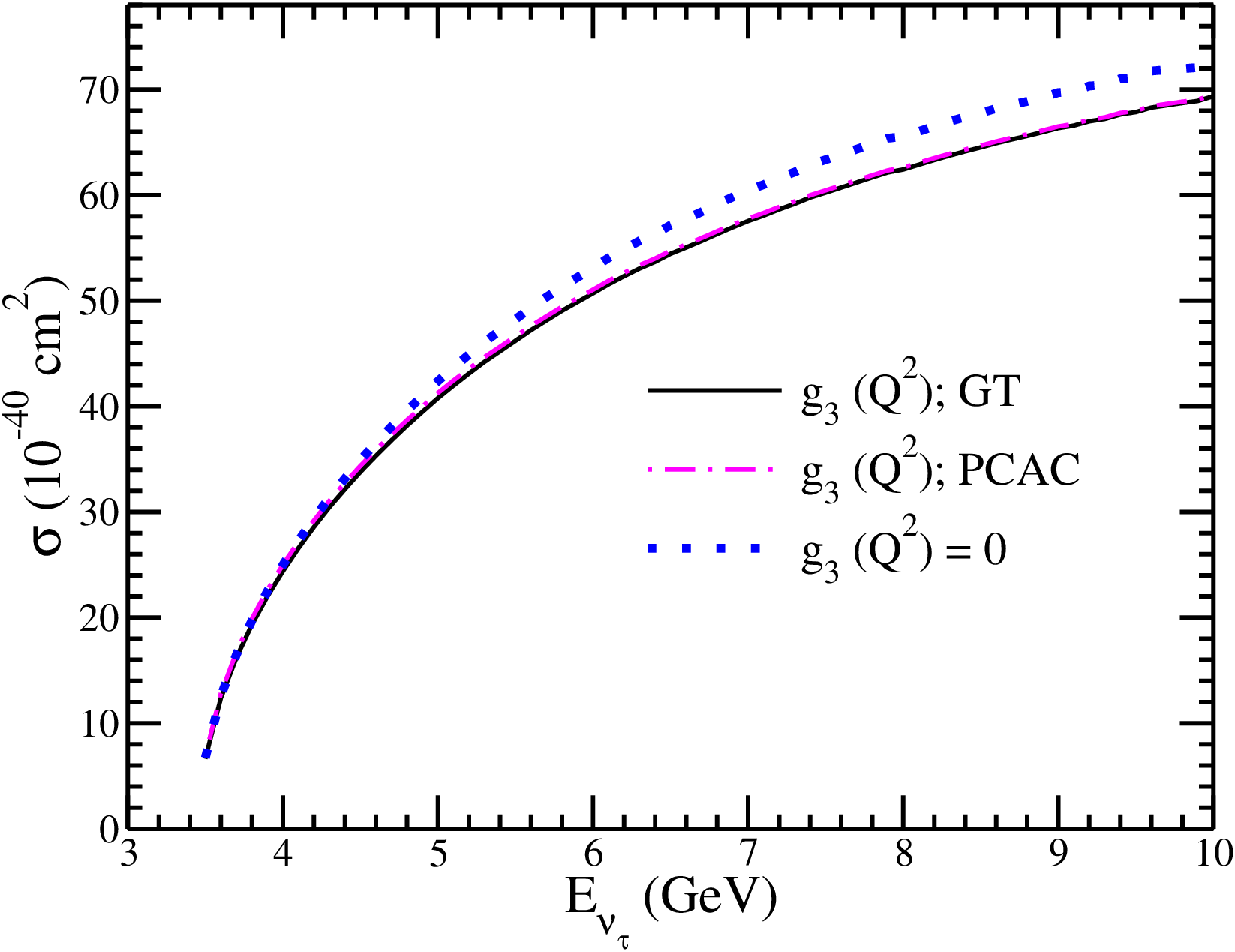}
\includegraphics[height=7.5cm,width=5.9cm]{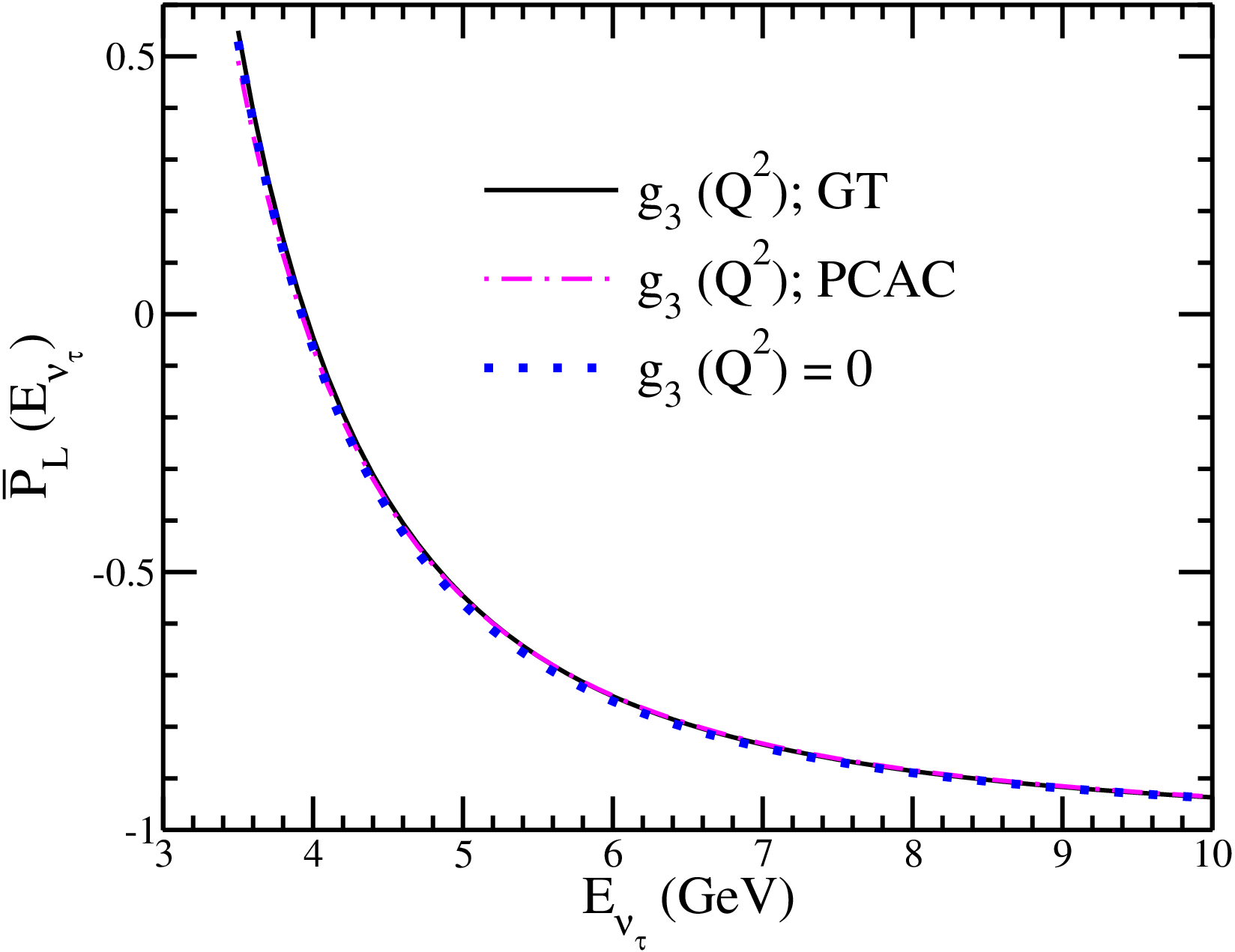}
\includegraphics[height=7.5cm,width=5.9cm]{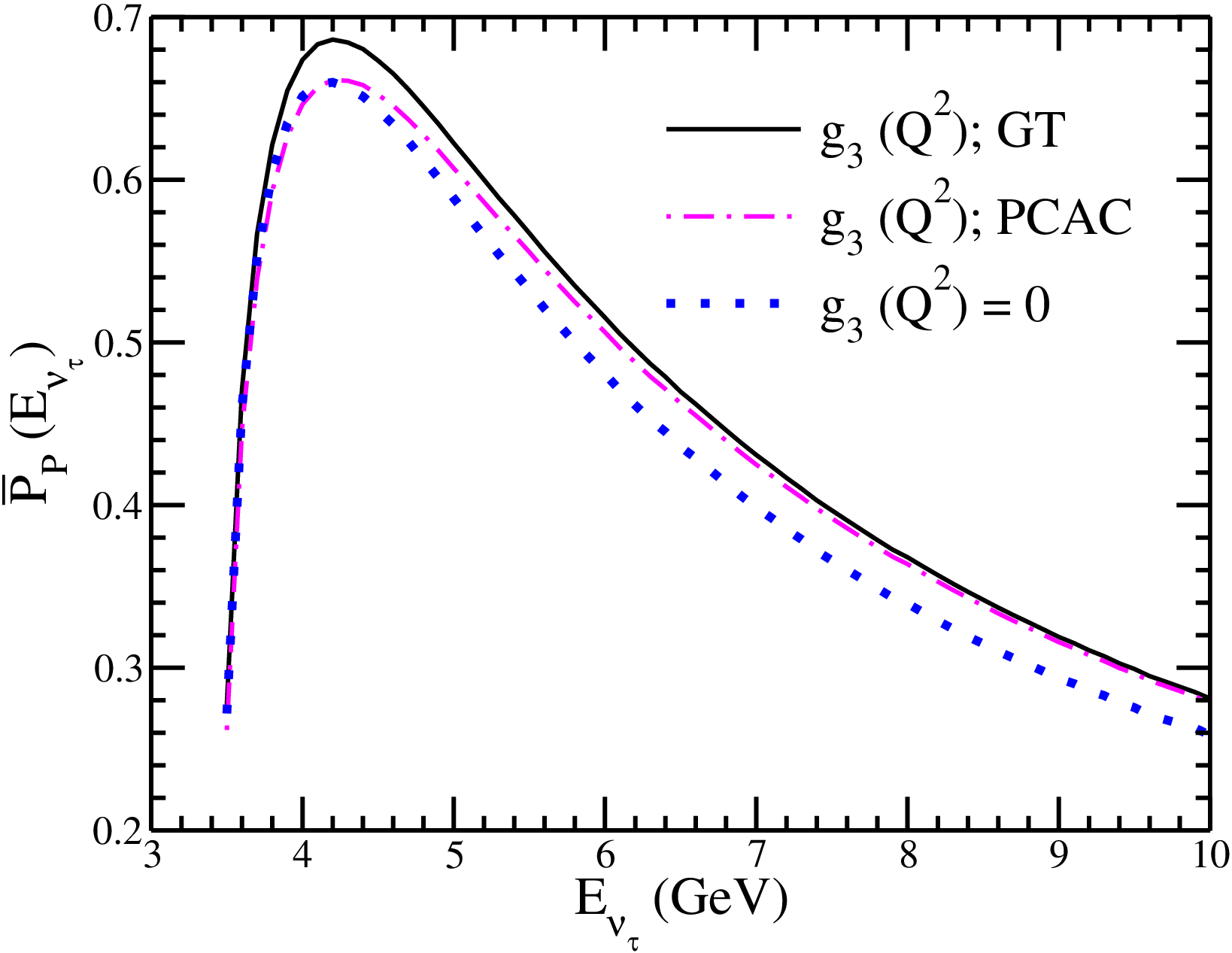}
\includegraphics[height=7.5cm,width=5.9cm]{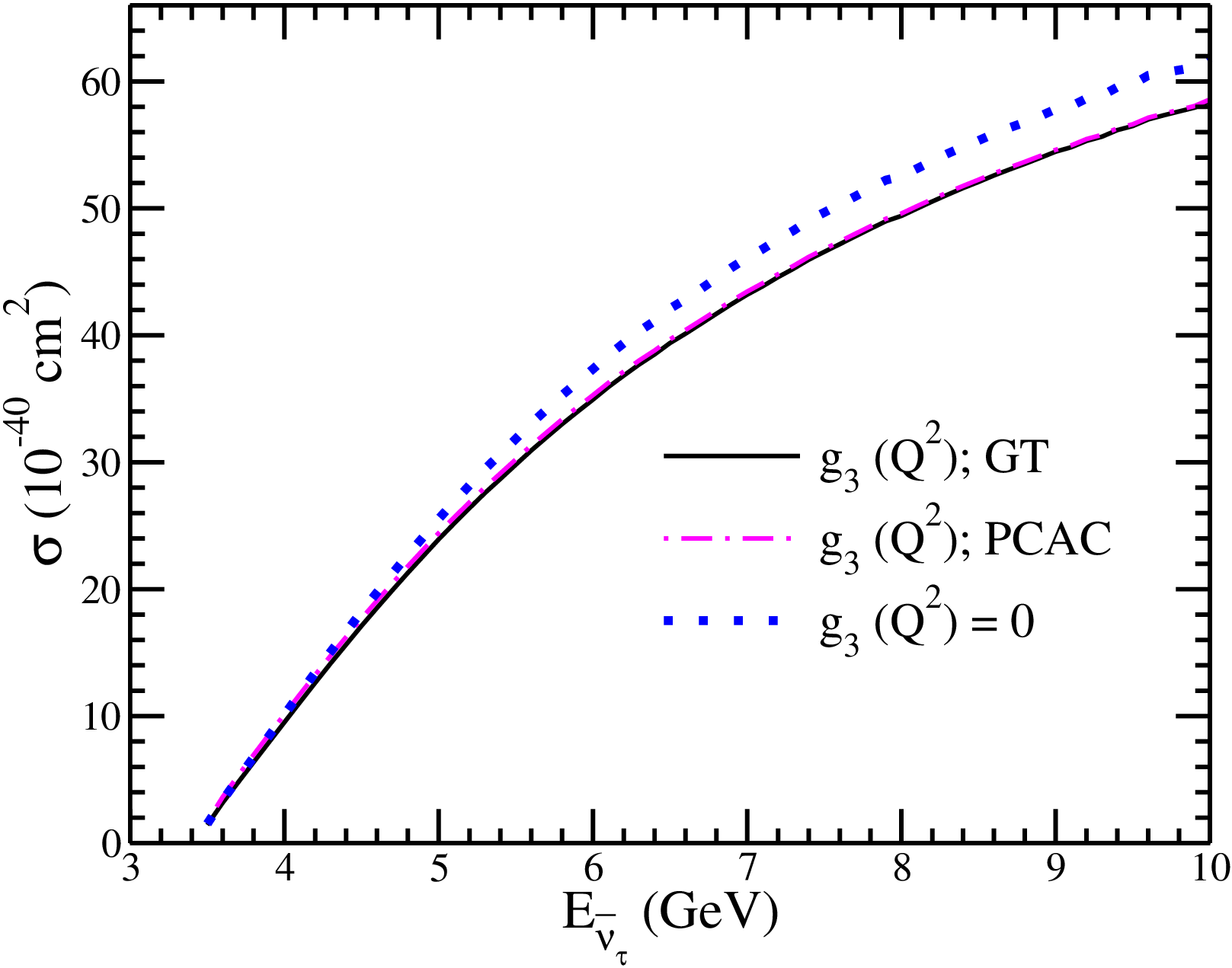}
\includegraphics[height=7.5cm,width=5.9cm]{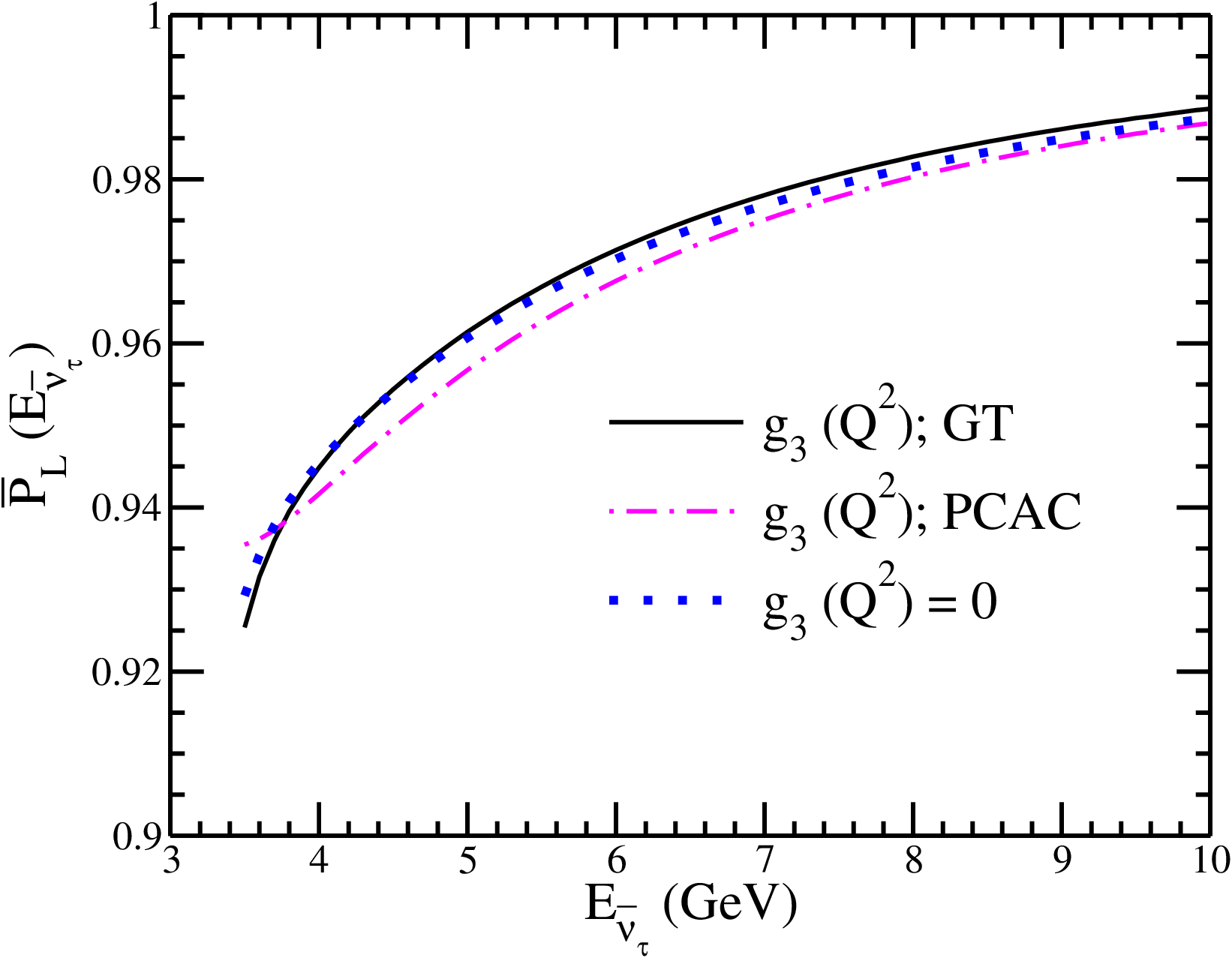}
\includegraphics[height=7.5cm,width=5.9cm]{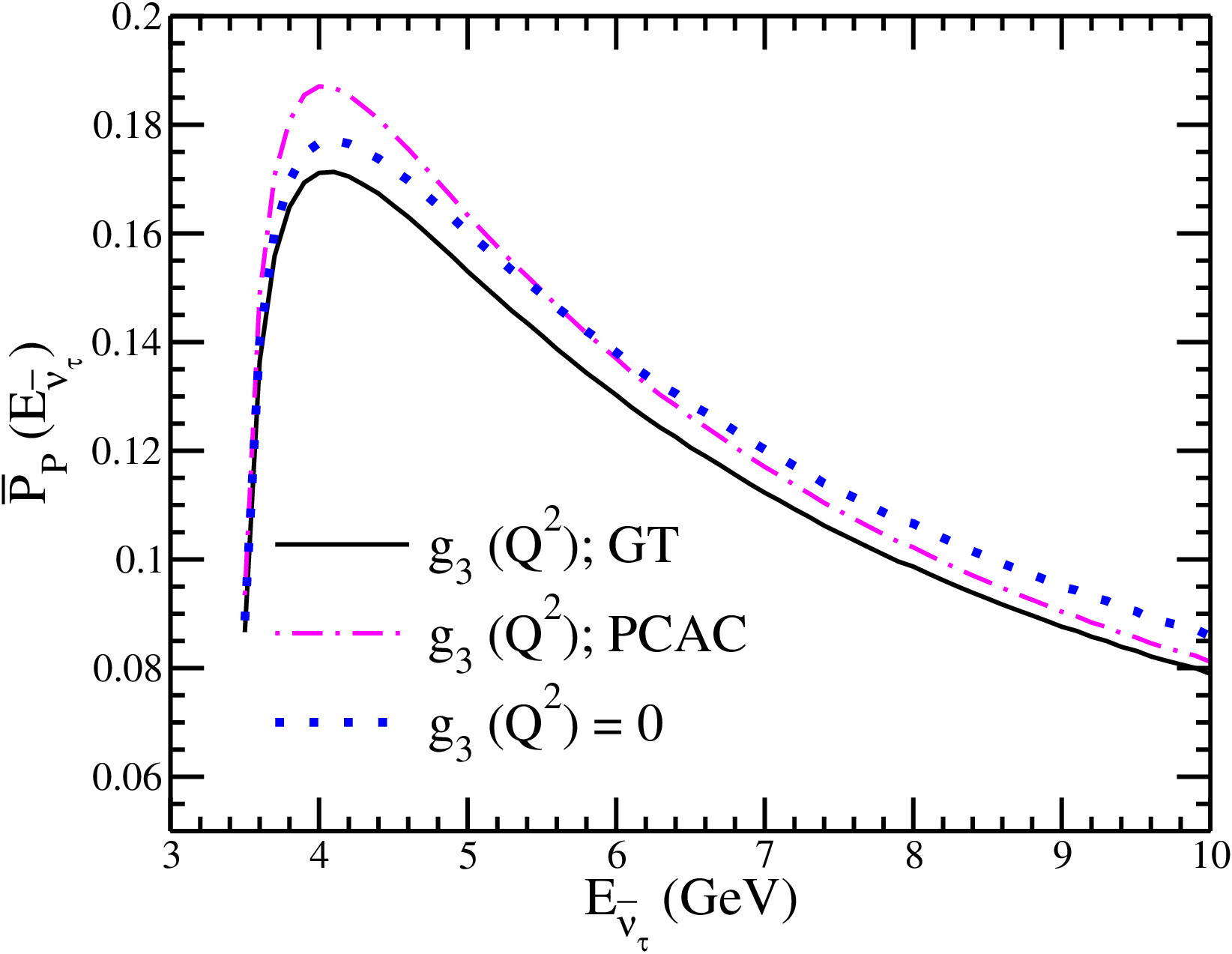}
\caption{(Top panel) Left to Right:  $\sigma$ vs $E_{\nu_{\tau}}$, $\overline{P}_{L} (E_{\nu_{\tau}})$ vs $E_{\nu_{\tau}}$,
 and $\overline{P}_{P} (E_{\nu_{\tau}})$ vs $E_{\nu_{\tau}}$ for the $\nu_{\tau} + n \rightarrow \tau^{-} + p$ process.
 (Bottom panel) Left to Right:  $\sigma$ vs $E_{\bar{\nu}_{\tau}}$, $\overline{P}_{L} (E_{\bar{\nu}_{\tau}})$ vs $E_{\bar{\nu}_{\tau}}$,
 and $\overline{P}_{P} (E_{\bar{\nu}_{\tau}})$ vs $E_{\bar{\nu}_{\tau}}$ for the $\bar{\nu}_{\tau} + p \rightarrow \tau^{+} + n$ process. The calculations have been performed using electric and magnetic Sachs form factors(Eqs.~(\ref{f1pn}) and (\ref{f2pn})) parameterized by Bradford {\it et al.}~\cite{Bradford:2006yz} and for the axial form factor (Eq.~(\ref{ga})), $M_{A} = 1.026$~GeV is used, with the different parameterizations of the pseudoscalar form factor {\it viz.}, using Goldberger-Treiman relation~(solid line), modified PCAC relation given by Schindler~\cite{Schindler:2006it}~(dashed-dotted line) and $g_{3} (Q^{2})=0$~(dotted line) as
discussed in section~\ref{W_FF} and the expressions are given in Eqs.~(\ref{g3}) and (\ref{fp1}).}\label{sigma_FP_nu}
\end{figure}

\begin{figure}
\centering
\includegraphics[height=7.5cm,width=5.9cm]{total_sigma_proton_g2r.eps}
\includegraphics[height=7.5cm,width=5.9cm]{total_Pl_proton_g2r.eps}
\includegraphics[height=7.5cm,width=5.9cm]{total_Pp_proton_g2r.eps}
\includegraphics[height=7.5cm,width=5.9cm]{total_sigma_neutron_g2r.eps}
\includegraphics[height=7.5cm,width=5.9cm]{total_Pl_neutron_g2r.eps}
\includegraphics[height=7.5cm,width=5.9cm]{total_Pp_neutron_g2r.eps}
\caption{(Top panel) Left to Right:  $\sigma$ vs $E_{\nu_{\tau}}$, $\overline{P}_{L} (E_{\nu_{\tau}})$ vs $E_{\nu_{\tau}}$,
 and $\overline{P}_{P} (E_{\nu_{\tau}})$ vs $E_{\nu_{\tau}}$ for the $\nu_{\tau} + n \rightarrow \tau^{-} + p$ process.
 (Bottom panel) Left to Right:  $\sigma$ vs $E_{\bar{\nu}_{\tau}}$, $\overline{P}_{L} (E_{\bar{\nu}_{\tau}})$ vs $E_{\bar{\nu}_{\tau}}$,
 and $\overline{P}_{P} (E_{\bar{\nu}_{\tau}})$ vs $E_{\bar{\nu}_{\tau}}$ for the $\bar{\nu}_{\tau} + p \rightarrow \tau^{+} + n$ process. The calculations have been performed using electric and magnetic Sachs form factors(Eqs.~(\ref{f1pn}) and (\ref{f2pn})) parameterized by Bradford {\it et al.}~\cite{Bradford:2006yz} and for the axial form factor (Eq.~(\ref{ga})), $M_{A} = 1.026$~GeV is used, with the different values of $g_{2}^{R} (0)$ {\it viz.} $g_{2}^{R} (0) = 0$~(solid line), 1~(dashed line) and $-1$~(double-dotted-dashed line) used in Eq.~(\ref{g2}).}\label{sigma_g2R_nu}
\end{figure}

\begin{figure}
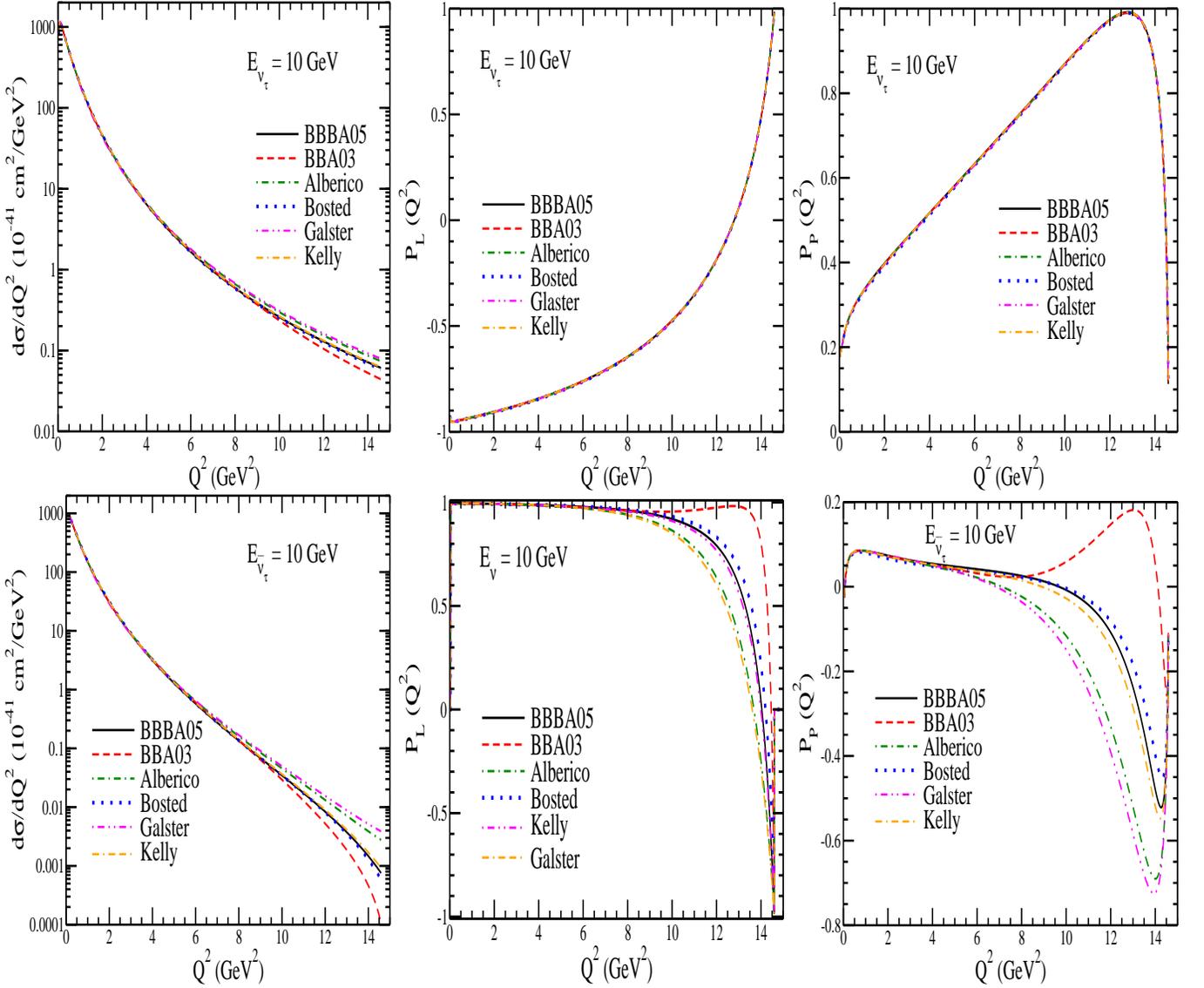

\centering
\includegraphics[height=7.5cm,width=5.9cm]{dsig_dq2_FF_variation_enu_10GeV_proton_polarized.eps}
\includegraphics[height=7.5cm,width=5.9cm]{Pl_q2_FF_variation_enu_10GeV_proton.eps}
\includegraphics[height=7.5cm,width=5.9cm]{Pp_q2_FF_variation_enu_10GeV_proton.eps}
\includegraphics[height=7.5cm,width=5.9cm]{dsig_dq2_FF_variation_enu_10GeV_neutron_polarized.eps}
\includegraphics[height=7.5cm,width=5.9cm]{Pl_q2_FF_variation_enu_10GeV_neutron.eps}
\includegraphics[height=7.5cm,width=5.9cm]{Pp_q2_FF_variation_enu_10GeV_neutron.eps}
\caption{$Q^2$-dependence of the differential scattering cross section $\frac{d\sigma}{dQ^{2}}$~(left panel), longitudinal polarization  $P_L (Q^2)$~(middle panel) and perpendicular polarization $P_P (Q^2)$~(right panel) for the processes 
$\nu_{\tau} + n \longrightarrow \tau^{-} + p$~(upper panel) and $\bar{\nu}_{\tau} + p \longrightarrow \tau^{+} + n$~(lower panel) when $\tau^-$~($\tau^{+}$) is polarized at  $E_{\nu_{\tau}, \bar{\nu}_{\tau}}$ 
= 10 GeV with $M_A$ = 1.026 GeV. Lines and points have the same meaning as in Fig.~\ref{sigma_VFF_nu}.}
\label{nutau_tau_pol_VFF_P}
\end{figure}

\begin{figure}
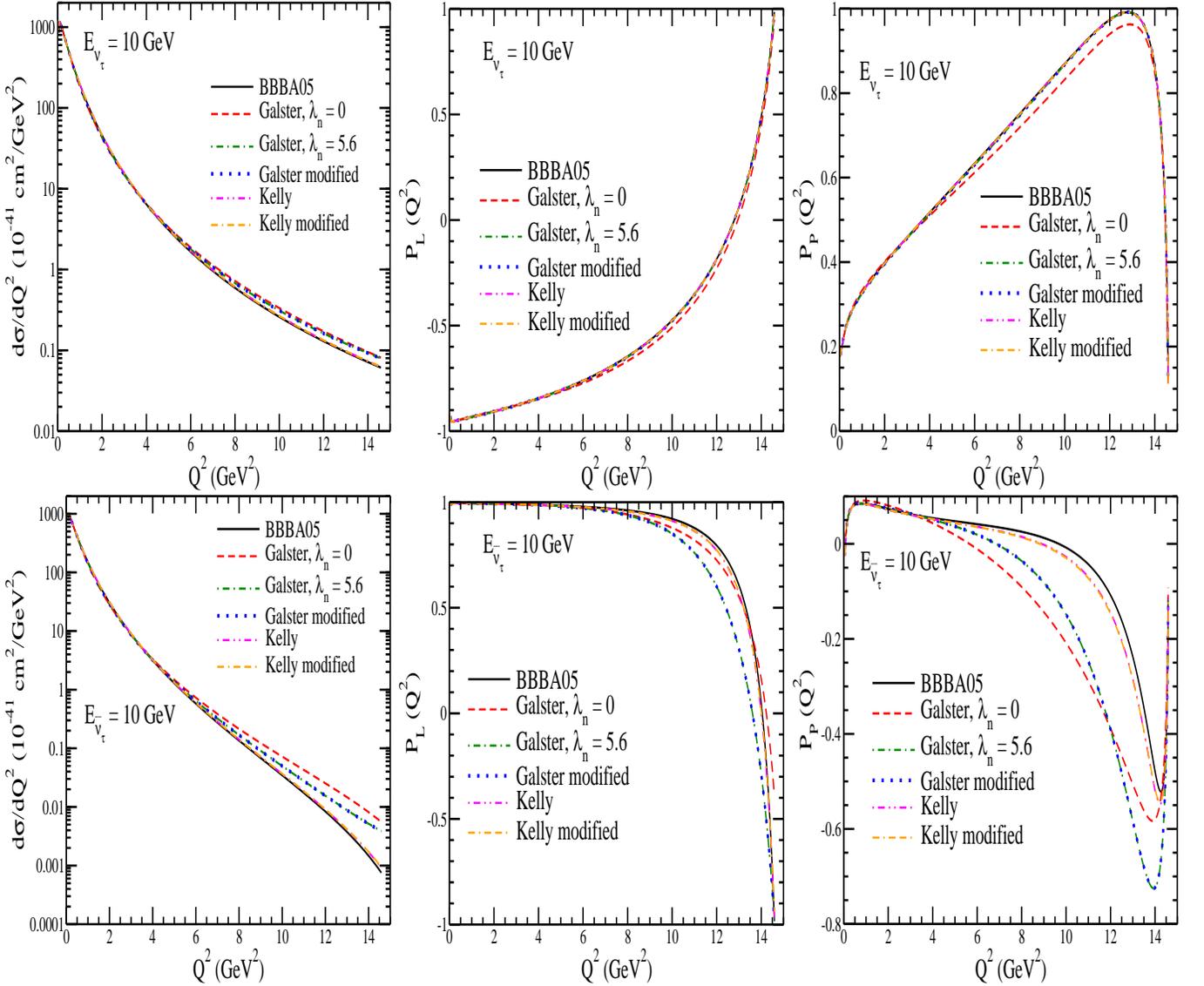

\centering
\includegraphics[height=7.5cm,width=5.9cm]{dsig_dq2_Gen_variation_enu_10GeV_proton_polarized.eps}
\includegraphics[height=7.5cm,width=5.9cm]{Pl_q2_Gen_variation_enu_10GeV_proton.eps}
\includegraphics[height=7.5cm,width=5.9cm]{Pp_q2_Gen_variation_enu_10GeV_proton.eps}
\includegraphics[height=7.5cm,width=5.9cm]{dsig_dq2_Gen_variation_enu_10GeV_neutron_polarized.eps}
\includegraphics[height=7.5cm,width=5.9cm]{Pl_q2_Gen_variation_enu_10GeV_neutron.eps}
\includegraphics[height=7.5cm,width=5.9cm]{Pp_q2_Gen_variation_enu_10GeV_neutron.eps}
\caption{$\frac{d\sigma}{dQ^{2}}$~(left panel),  $P_L (Q^2)$~(middle panel) and $P_P (Q^2)$~(right panel) versus $Q^2$ for the processes 
$\nu_{\tau} + n \longrightarrow \tau^{-} + p$~(upper panel) and $\bar{\nu}_{\tau} + p \longrightarrow \tau^{+} + n$~(lower panel) when $\tau^-~(\tau^{+})$ is polarized at  $E_{\nu_{\tau}}$ 
= 10 GeV with $M_A$ = 1.026 GeV. Lines and points have the same meaning as in Fig.~\ref{sigma_Gen_nu}.}
\label{nutau_tau_pol_Gen_P}
\end{figure}

In Fig.~\ref{sigma_VFF_nu}, we have presented the results for the total cross section~($\sigma$) as well as the average longitudinal ~($\overline{P}_{L} (E_{\nu_{\tau}})$) and perpendicular ~($\overline{P}_{P} (E_{\nu_{\tau}})$) polarizations for the process $\nu_{\tau} + n \rightarrow \tau^{-} + p$ in the top panel, and in the bottom panel, the corresponding results are presented for the process
 $\bar{\nu}_{\tau} + p \rightarrow \tau^{+} + n$. 
  In this figure, we have studied the effect of the vector form factors~(Eqs.~\ref{f12pn}, \ref{f1pn} and \ref{f2pn}) on the total cross section and average polarizations by taking into account the different parameterizations for the Sachs' electric and magnetic form factors of the nucleon given by Bradford {\it et al.}, known as BBBA05~\cite{Bradford:2006yz}, Budd {\it et al.}, known as BBA03~\cite{Budd:2004bp}, Bosted~\cite{Bosted:1994tm}, Alberico {\it et al.}~\cite{Alberico:2008sz}, Kelly~\cite{Kelly:2004hm} and Galster {\it et al.}~\cite{Galster:1971kv} which are given in the Appendix. In the case of $\nu_{\tau} + n \rightarrow \tau^{-} + p$, the results for the 
  total cross section using the different parameterizations of the electric and magnetic nucleon form factors are almost consistent 
  with each other, except for the results obtained using the parameterization of Galster {\it et al.}~\cite{Galster:1971kv} which 
  are slightly lower~(1--2$\%$) than the results of the other parameterizations for $E_{\nu_{\tau}} \le 6$~GeV. However, beyond $E_{\nu_{\tau}}=6$~GeV, the results from all the other parameterizations except the parameterization of Galster {\it et al.}, are  consistent up to 10 GeV whereas the results with the parameterization of Galster {\it et al.}~\cite{Galster:1971kv} are higher. At $E_{\nu_{\tau}}=10$~GeV, the results obtained with the 
  parameterization of Galster {\it et al.}~\cite{Galster:1971kv} are $\sim$ 7\% larger than the results obtained using the parameterizations of other works~\cite{Bradford:2006yz,Budd:2004bp,Bosted:1994tm,Alberico:2008sz,Kelly:2004hm}. In the case of the average polarizations for the process $\nu_{\tau} + n \rightarrow \tau^{-} + p$, the results obtained with the different parameterizations of the Sachs's form factors are consistent with one another except for the parameterization of Galster {\it et al.}~\cite{Galster:1971kv} throughout the range of $E_{\nu_{\tau}}$. The results obtained with Galster's parameterization are smaller than the results obtained with other parameterizations for both $\overline{P}_{L} (E_{\nu_{\tau}})$ and $\overline{P}_{P} (E_{\nu_{\tau}})$. 
  
  In the case of $\bar{\nu}_{\tau} + p \rightarrow \tau^{+} + n$ scattering process, unlike the neutrino induced process, the results obtained with Galster {\it et al.}~\cite{Galster:1971kv} are higher than the results obtained with other parameterizations in the full range of $E_{\bar{\nu}_{\tau}}$ {\it i.e.}, from threshold up to 10 GeV, while the results from all the other parameterizations are consistent among themselves. For example, at $E_{\bar{\nu}_{\tau}} =5~(10)$ GeV, the results obtained with Galster {\it et al.}~\cite{Galster:1971kv} are $\sim$ 30\%~(23\%) higher than the results obtained with BBBA05~\cite{Bradford:2006yz}. In the case of average polarizations, $\overline{P}_{L} (E_{\bar{\nu}_{\tau}})$ shows similar trend as we have observed in the case of $\nu_{\tau} + n \rightarrow \tau^{-} + p$, {\it i.e.}, the results obtained with the parameterization of Galster {\it et al.}~\cite{Galster:1971kv} are lower than the results of the other parameterizations which are consistent among themselves, while the results of $\overline{P}_{P} (E_{\bar{\nu}_{\tau}})$ using Galster {\it et al.}~\cite{Galster:1971kv} parameterization become almost zero. This is in remarkable difference from the results obtained using the parameterizations from other works~\cite{Bradford:2006yz,Budd:2004bp,Bosted:1994tm,Alberico:2008sz,Kelly:2004hm}.

To see the dependence of the electric form factor of neutron~($G_{E}^{n} (Q^2)$) on the total cross section and average polarizations for the processes $\nu_{\tau} + n \rightarrow \tau^{-} + p$ and $\bar{\nu}_{\tau} + p \rightarrow \tau^{+} + n$, we have shown 
 in Fig.~\ref{sigma_Gen_nu},
 the results for $\sigma$, $\overline{P}_{L} (E_{\nu_{\tau}( \bar{\nu}_{\tau})})$ and $\overline{P}_{P} (E_{\nu_{\tau}(\bar{\nu}_{\tau})})$ as a function of $E_{\nu_{\tau} ( \bar{\nu}_{\tau})}$, by taking into account the different parameterizations of $G_{E}^{n} (Q^2)$ available in the literature {\it viz.}, the parameterization given by Galster {\it et al.}~\cite{Galster:1971kv}, its modification by Platchkov {\it et al.}~\cite{Platchkov:1989ch}, the parameterization given by Kelly~\cite{Kelly:2004hm} and its modification by Punjabi {\it et al.}~\cite{Punjabi:2015bba}. For comparison, we have also presented the results using BBBA05 parameterization~\cite{Bradford:2006yz}. It may be observed from the figure that in the case of $\sigma$ for the process $\nu_{\tau} + n \rightarrow \tau^{-} + p$, the results obtained with BBBA05~\cite{Bradford:2006yz}, Kelly~\cite{Kelly:2004hm} and its modification~\cite{Punjabi:2015bba} are consistent with one another from threshold up to $E_{\nu_{\tau}} =10$~GeV, while the results obtained with Galster parameterization and its modification are almost comparable, but are higher than the results of BBBA05~\cite{Bradford:2006yz}. In the case of $\overline{P}_{L} (E_{\nu_{\tau}})$ and $\overline{P}_{P} (E_{\nu_{\tau}})$ for $\nu_{\tau} + n \rightarrow \tau^{-} + p$, the results show similar trend as observed in the case of Sachs' form factor variation~(Fig.~\ref{sigma_VFF_nu}), {\it i.e.}, the results obtained with the parameterization of BBBA05~\cite{Bradford:2006yz}, Kelly~\cite{Kelly:2004hm} and its modification~\cite{Punjabi:2015bba} are in a very good agreement with each other but are higher than the results of Galster {\it et al.} parameterization~\cite{Galster:1971kv} and its modification~\cite{Platchkov:1989ch}. For the process $\bar{\nu}_{\tau} + p \rightarrow \tau^{+} + n$, the results obtained for the total cross section as well as the average polarizations show similar trend as observed in the case of $\nu_{\tau}$ induced process, {\it i.e.}, the results obtained with the parameterization of Galster {\it et al.}~\cite{Galster:1971kv} and its modification~\cite{Platchkov:1989ch} are higher than the results of BBBA05~\cite{Bradford:2006yz} parameterization for the total cross section, while in the case of $\overline{P}_{L} (E_{\bar{\nu}_{\tau}})$ the results obtained with the parameterization of Galster {\it et al.}~\cite{Galster:1971kv} and its modification~\cite{Platchkov:1989ch} are lower than the results of BBBA05~\cite{Bradford:2006yz} whereas
 in the case of $\overline{P}_{P} (E_{\bar{\nu}_{\tau}})$, the results using Galster {\it et al.}~\cite{Galster:1971kv} parameterization and its modification~\cite{Platchkov:1989ch} are smaller than the results obtained using 
 the parameterizations of BBBA05~\cite{Bradford:2006yz}, Kelly~\cite{Kelly:2004hm} and its modification~\cite{Punjabi:2015bba} which are significantly higher and non-negligible.

To study the effect of the axial dipole mass $M_{A}$ on $\sigma$ as well as on $\overline{P}_{L} (E_{\nu_{\tau}})$ and $\overline{P}_{P} (E_{\nu_{\tau}})$, we have presented in Fig.~\ref{sigma_MA_nu} the results for $\nu_{\tau} + n \rightarrow \tau^{-} + p$ process in the top panel and in the bottom panel, the results are presented for the 
 $\bar{\nu}_{\tau} + p \rightarrow \tau^{+} + n$ scattering process.
   We have varied $M_{A}$ in the range 0.9--1.2 GeV which has been suggested as the range of the possible values of $M_A$ in the different works available in the literature~\cite{Katori:2016yel}. From the figure, it may be observed that with the increase in 
   $M_{A}$,  the strength of the cross section increases and this increase is significant even near the threshold region. For example, 
   by increasing $M_{A}$ by 20\% from the world average value, the cross section for the process $\nu_{\tau} + n \rightarrow \tau^{-} + p$ at $E_{\nu_{\tau}} =3.5$, 5 and 10 GeV, respectively, increases by about 40\%, 28\% and 20\% while for the process $\bar{\nu}_{\tau} + p \rightarrow \tau^{+} + n$, $\sigma$ increases by $\sim$ 60\%, 25\% and 19\%, respectively, at $E_{\bar{\nu}_{\tau}} =3.5$, 5 and 10 GeV.
   Similarly, by decreasing the value of $M_{A}$ by 10\% from the world average value {i.e.}, $M_{A}=1.026$~GeV, the cross section for the process $\nu_{\tau} + n \rightarrow \tau^{-} + p$ decreases by $\sim$ 24\%, 17\% and 15\% at $E_{\nu_{\tau}} =3.5$, 5 and 10 GeV, respectively, while for the process $\bar{\nu}_{\tau} + p \rightarrow \tau^{+} + n$, the decrease in $\sigma$ is about 22\%, 15\% and 12\% at $E_{\bar{\nu}_{\tau}} =3.5$, 5 and 10 GeV, respectively.  Although, there exists large dependence of $M_{A}$ on the total cross section, the average longitudinal polarization $\overline{P}_{L} (E_{\nu_{\tau}(\bar{\nu}_{\tau})})$ is almost insensitive to the variation in the value of $M_{A}$ for the neutrino as well as antineutrino induced quasielastic scattering processes and also for $\overline{P}_{P} (E_{\nu_{\tau}})$ in the case 
   of neutrino induced process, while $\overline{P}_{P} (E_{\bar{\nu}_{\tau}})$ in the case of 
   antineutrino induced process shows significant $M_{A}$ dependence for the lower values of $M_A$ at $E_{\bar{\nu}_{\tau}}=4$ and 5 GeV, for example, by decreasing $M_{A}$ by 10\%, decreases $\overline{P}_{P} (E_{\bar{\nu}_{\tau}})$ by $\sim$ 60\% and 35\%, respectively, and this difference in the values $\overline{P}_{P} (E_{\bar{\nu}_{\tau}})$ gradually becomes smaller with the increase in $M_A$. 
   
   This can be compared with the corresponding results in the case of $\nu_{\mu} (\bar{\nu}_{\mu}) - N$ scattering. In the case of $\nu_{\mu} (\bar{\nu}_{\mu})$ induced quasielastic scattering, the threshold for the muon production is $\sim$ 0.1 GeV, while in the case the $\nu_{\tau} (\bar{\nu}_{\tau})$ induced processes, the threshold is $\sim$ 3.5 GeV. In the case of $\nu_{\mu}$ induced process, increasing $M_A$ by 20\%, increases the cross section at $E_{\nu_{\mu}}=0.2$~GeV~(near the threshold) by 3\%, while at $E_{\nu_{\mu}} =$ 1, 2 and 3 GeV, $\sigma$ increases by $\sim$ 15\%, 18\% and 18\%, respectively~\cite{Fatima:2018tzs}.
    In contrast, we have observed~(Fig.~\ref{sigma_MA_nu}) the increase in $\sigma$ to be 48\%, 28\% and 20\% at $E_{\nu_{\tau}}=3.5$, 5 and 10 GeV, respectively, when $M_{A}$ is increased by 20\% from the world average value in the case of $\nu_{\tau} - N$ scattering. It may be pointed that in the case of $\nu_{\mu}$ induced processes, the percentage increment in $\sigma$ obtained using $M_{A}=1.2$~GeV, increases with increase in $E_{\nu_{\mu}}$, while in the case of $\nu_{\tau}$ induced processes, the percentage increment in $\sigma$ decreases with increase in the energy of the incoming neutrino. Similarly, in the case of $\bar{\nu}_{\tau}$ induced quasielastic scattering process, we observe a similar trend as has been observed in the case of $\nu_{\tau}$ scattering, but quantitatively the percentage increment in the cross sections are smaller in the case of ${\bar\nu}_{\tau} - N$ scattering.

To study the effect of pseudoscalar form factor on the total cross section $\sigma$, and the polarization observables 
 $\overline{P}_{L} (E_{{\nu}_{\tau}}, E_{\bar{\nu}_{\tau}})$ and $\overline{P}_{P} (E_{{\nu}_{\tau}}, E_{\bar{\nu}_{\tau}})$ for the processes $\nu_{\tau} + n \rightarrow \tau^{-} + p$ and $\bar{\nu}_{\tau} + p \rightarrow \tau^{+} + n$, we have used two parameterizations
 as given in Eqs.~(\ref{g3}) and (\ref{fp1}). The numerical results are presented in Fig.~\ref{sigma_FP_nu}.
 It may be observed from the figure that the results obtained using PCAC + GT and PCAC + modified GT are consistent for $\sigma$ for
 both the neutrino as well as antineutrino induced processes and tend to decrease the cross sections by $\sim$ 3\% and 6\%, respectively, for these processes at $E_{\nu_{\tau} (\bar{\nu}_{\tau})}=10$ GeV. This decrease in the results due to the inclusion of $g_{3} (Q^2)$ implies that in the case of $\nu_{\tau}(\bar{\nu}_{\tau})-N$ scattering, there is non-negligible contribution from the pseudoscalar form factor specially at higher energies. For both the processes, $\overline{P}_{L} (E_{\nu_{\tau}, \bar{\nu}_{\tau}})$ is almost insensitive to the different parameterizations of $g_{3} (Q^2)$ while $\overline{P}_{P} (E_{\nu_{\tau}, \bar{\nu}_{\tau}})$ shows  some dependence and the nature of dependence is different for the neutrino and the antineutrino induced processes as 
 shown in these figures.

To observe the dependence of the second class current form factor, we have varied $g_{2}^{R} (0)$ used in Eq.~(\ref{g2}), and studied its effect on the total cross section and average polarizations. In Fig.~\ref{sigma_g2R_nu}, we have presented the results for $\sigma$, $\overline{P}_{L} (E_{\nu_{\tau}(\bar{\nu}_{\tau})})$ and $\overline{P}_{P} (E_{\nu_{\tau}(\bar{\nu}_{\tau})})$ as a function of neutrino~(top panel)/antineutrino~(bottom panel) energies by taking $g_{2}^{R} (0)=0$ and $\pm 1$. There is some information about the value of $g_{2}^{R} (0)$ from the muon capture and quasielastic scattering but it is not conclusive. The value of the dipole mass $M_2$  
 can be determined from the analysis of the quasielastic neutrino scattering. In the absence of such analyses $M_2$ is taken to be  equal to $M_A$ in the dipole parameterization of the 
form factor $g_{2} (Q^2)$ and a non-zero value of $g_{2}^{R} (0)$ is chosen for the purpose of illustrating the quantitative effect of the second class currents in $\nu_\tau - N$ scattering. 
 It may be observed from the figure that in the case of $\sigma$, for both the processes $\nu_{\tau} + n \rightarrow \tau^{-} + p$ and $\bar{\nu}_{\tau} + p \rightarrow \tau^{+} + n$, the results obtained with $g_{2}^{R} (0) =-1$ are slightly lower~(1 -- 2$\%$) than the results obtained with $g_{2}^{R} (0) =0$ in the range of $E_{\nu_{\tau}, \bar{\nu}_{\tau}}$ from threshold up to 10 GeV, while the results obtained with $g_{2}^{R} (0) =+1$, are higher from the results obtained with $g_{2}^{R} (0) =0$  and the difference decreases with the increase in energy. For example, at $E_{\nu_{\tau} (\bar{\nu}_{\tau})}=5$~GeV, the results obtained with $g_{2}^{R} (0) = +1$ are higher by about 18~(30)$\%$ from the results of $g_{2}^{R} (0)=0$, while at 
 10 GeV, this difference becomes 10~(12)$\%$ for the neutrino~(antineutrino)  induced processes.
  In the case of $\overline{P}_{L} (E_{\nu_{\tau}, \bar{\nu}_{\tau}})$, there is a slight variation due to the change in the value of $g_{2}^{R} (0)$ for  the neutrino induced process, while for the antineutrino induced process, this difference is larger at lower antineutrino energies which gradually becomes smaller with the increase in energy. For $\overline{P}_{P} (E_{\nu_{\tau} (\bar{\nu}_{\tau})})$, the results for both the neutrino as well as antineutrino induced processes show dependence on the choice of $g_{2}^{R} (0)$, while the nature of dependence is different.
 In the case of $\nu_\tau$ induced reaction, in the peak region, the results are $\sim$ 20$\%$ smaller for $g_{2}^{R} (0)=+1$ from the results obtained with $g_{2}^{R} (0)=0$, while using $g_{2}^{R} (0)=-1$ the results are 18$\%$ higher than the results obtained using $g_{2}^{R} (0) =0$. However, in the case of $\bar{\nu}_{\tau}$ induced processes, the results obtained with $g_{2}^{R} (0) = \pm 1$ are lower than the results obtained with $g_{2}^{R} (0)=0$ in the region of threshold up to $E_{\bar{\nu}_{\tau}}=6$ GeV. In the threshold energy region, the results with $g_{2}^{R} (0) = 0$ and $g_{2}^{R} (0) = - 1$ are close by while at high energies $E_{\bar{\nu}_{\tau}} (> 5$~GeV), the difference in the results of $g_{2}^{R} (0) = 0$ and $g_{2}^{R} (0) = + 1$ is quite small.

\section{Results and discussion: Differential scattering cross section and $Q^{2}$-dependent polarization observables}\label{cross_section}
To see the dependence of $\frac{d\sigma}{dQ^2}$,  $P_{L} (Q^{2})$ and $P_{P} (Q^{2})$ on the nucleon vector form factors, we have presented in 
Fig.~\ref{nutau_tau_pol_VFF_P} the numerical results for the production of polarized $\tau^-$ in the process $\nu_{\tau} + n \rightarrow \tau^{-} + p$ as well as for the polarized $\tau^{+}$ in the process 
$\bar{\nu}_{\tau} + p \rightarrow \tau^{+} + n$ by using the different parameterizations of Sachs' electric and magnetic form factors given in Appendix-A. In the case of differential scattering cross section for both the processes, at lower energies of the incoming neutrino and antineutrino, there is hardly any variation~(not shown here) due to the different parameterizations of the Sachs's electric and magnetic form factors. Therefore, the results have been presented only at $E_{\nu_{\tau} (\bar{\nu}_{\tau})}=$ 10 GeV for the purpose of illustrating the main features. At $E_{\nu_{\tau}~(\bar{\nu}_{\tau})}=10$ GeV, the differential scattering cross section obtained using the different parameterizations of vector form factors are different only in the high $Q^2$ region where the parameterizations of BBBA05~\cite{Bradford:2006yz}, Kelly~\cite{Kelly:2004hm} and Bosted~\cite{Bosted:1994tm} are consistent with one another while the results obtained with the parameterizations of Alberico~\cite{Alberico:2008sz} and Galster~\cite{Galster:1971kv} are higher than the one obtained with BBBA05~\cite{Bradford:2006yz}. Furthermore, this difference in the differential cross section for the various parameterizations is more pronounced in the case of $\bar{\nu}_{\tau}$ induced process than in $\nu_{\tau}$ induced process. In the case of $\nu_{\tau}$ induced process, there is almost no sensitivity of the polarization components $P_{L} (Q^2)$ and $P_{P} (Q^2)$ when the different parameterizations for the Sachs' electric and magnetic form factors are used. However, in the case of $\bar{\nu}_{\tau}$ induced process, we observe large dependence of $P_{L} (Q^{2})$ and 
$P_{P} (Q^{2})$ on the different parameterizations of Sachs' electric and magnetic form factors of nucleon for $Q^{2} 
\ge 8$ GeV$^{2}$ in the case of $P_{L} (Q^{2})$ and for $Q^{2} \ge 6$ GeV$^{2}$ in the case of $P_{P} (Q^{2})$. 

Similarly, in Fig.~\ref{nutau_tau_pol_Gen_P}, we have presented the results for the  polarized $\tau^{-}$ produced in the 
reaction ${\nu}_{\tau} + n \rightarrow \tau^{-} + p$ and for the polarized $\tau^{+}$ produced in the 
reaction $\bar{\nu}_{\tau} + p \rightarrow \tau^{+} + n$ for $\frac{d\sigma}{dQ^2}$, $P_{L} (Q^{2})$ and $P_{P} (Q^{2})$ as a 
function of $Q^{2}$ using the different parameterization of $G_{E}^{n} (Q^{2})$ given by BBBA05~\cite{Bradford:2006yz}, Galster {\it et al.}~\cite{Galster:1971kv} using $\lambda_{n} =0$ and 5.6, modified Galster parameterization given by Platchkov {\it et al.}~\cite{Platchkov:1989ch}, Kelly~\cite{Kelly:2004hm} and modified Kelly parameterization given by Punjabi {\it et al.}~\cite{Punjabi:2015bba} at $E_{\bar{\nu}_{\tau}}=$ 10 GeV. It may be observed from the figure that in the case of $\frac{d\sigma}{dQ^2}$ for both the processes, the results obtained with BBBA05~\cite{Bradford:2006yz}, Kelly~\cite{Kelly:2004hm} and its modification by Punjabi~\cite{Punjabi:2015bba} are consistent with each other while the results obtained using the parameterization of Galster {\it et al.}~\cite{Galster:1971kv} with $\lambda_{n} =5.6$ and its modification by Platchkov {\it et al.}~\cite{Platchkov:1989ch} are consistent with one another but higher than the results with BBBA05~\cite{Bradford:2006yz}. This effect is more pronounced in the case of $\bar{\nu}_{\tau}$ induced process than in $\nu_{\tau}$ induced process. The polarization components 
 $P_{L} (Q^{2})$ and $P_{P} (Q^{2})$ in the case of $\nu_{\tau}$ induced process,  are almost insensitive to the different parameterizations of $G_{E}^{n} (Q^{2})$ while in the case of $\bar{\nu}_{\tau}$ induced process, there is a little dependence of $P_{L} (Q^{2})$ on the different parameterizations of $G_{E}^{n} 
(Q^{2})$, whereas we observe significant variation in $P_{P} (Q^{2})$ for 4 GeV$^{2} \le Q^{2} \le 14$ GeV$^{2}$. Moreover, it is worth mentioning that the results obtained with Kelly~\cite{Kelly:2004hm} parameterization and its modification~\cite{Platchkov:1989ch} as well as with Galster~\cite{Galster:1971kv} using $\lambda =5.6$ and the 
modification of Galster's parameterization given by Platchkov {\it et al.}~\cite{Platchkov:1989ch} give almost the same results for both 
$P_{L} (Q^{2})$ and $P_{P} (Q^{2})$ at all the values of $E_{\bar{\nu}_{\tau}}$ and $Q^{2}$ considered in this work. 

\begin{figure}
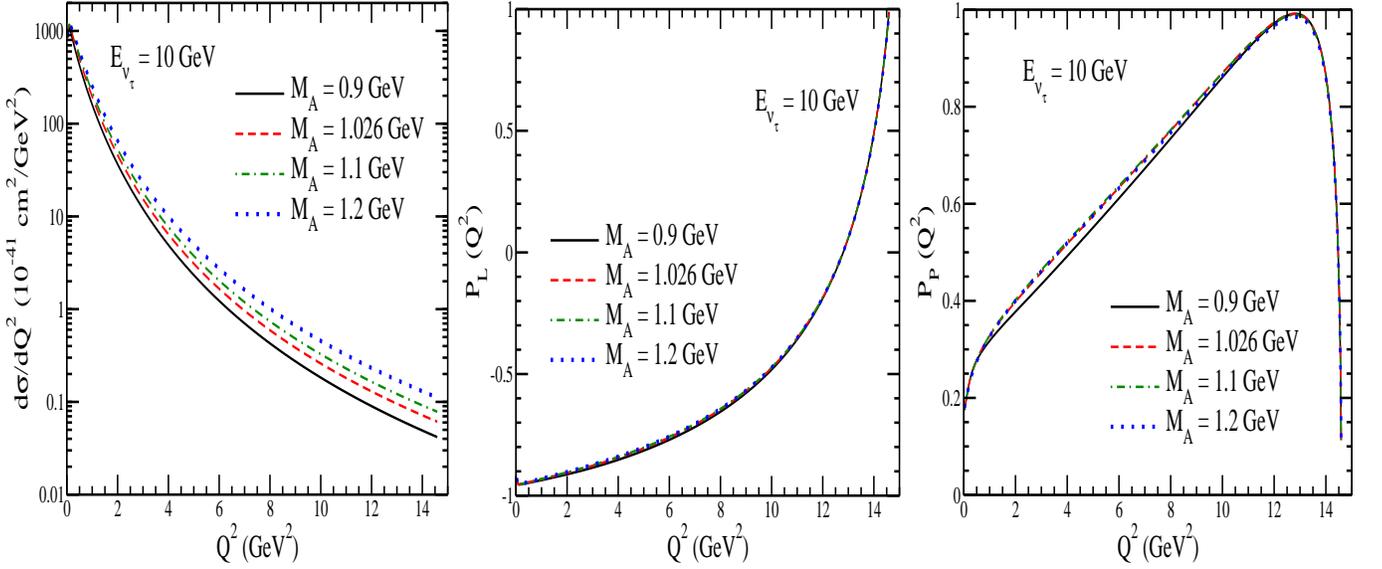

\centering
\includegraphics[height=7.5cm,width=5.9cm]{dsig_dq2_Ma_variation_enu_10GeV_proton_polarized.eps} 
\includegraphics[height=7.5cm,width=5.9cm]{Pl_q2_MA_variation_enu_10GeV_proton.eps}
\includegraphics[height=7.5cm,width=5.9cm]{Pp_q2_MA_variation_enu_10GeV_proton.eps}
\caption{$\frac{d\sigma}{dQ^{2}}$~(left panel), $P_L (Q^2)$~(middle panel) and $P_P (Q^2)$~(right panel) versus $Q^2$ for the process 
${\nu_{\tau}} + n \longrightarrow \tau^- + p$ when $\tau^-$ is polarized at $E_{\nu_{\tau}}$ = 10 GeV. Lines and points have the same meaning as in Fig.~\ref{sigma_MA_nu}.}\label{nutau_tau_pol_MA_P}
\end{figure} 

\begin{figure}
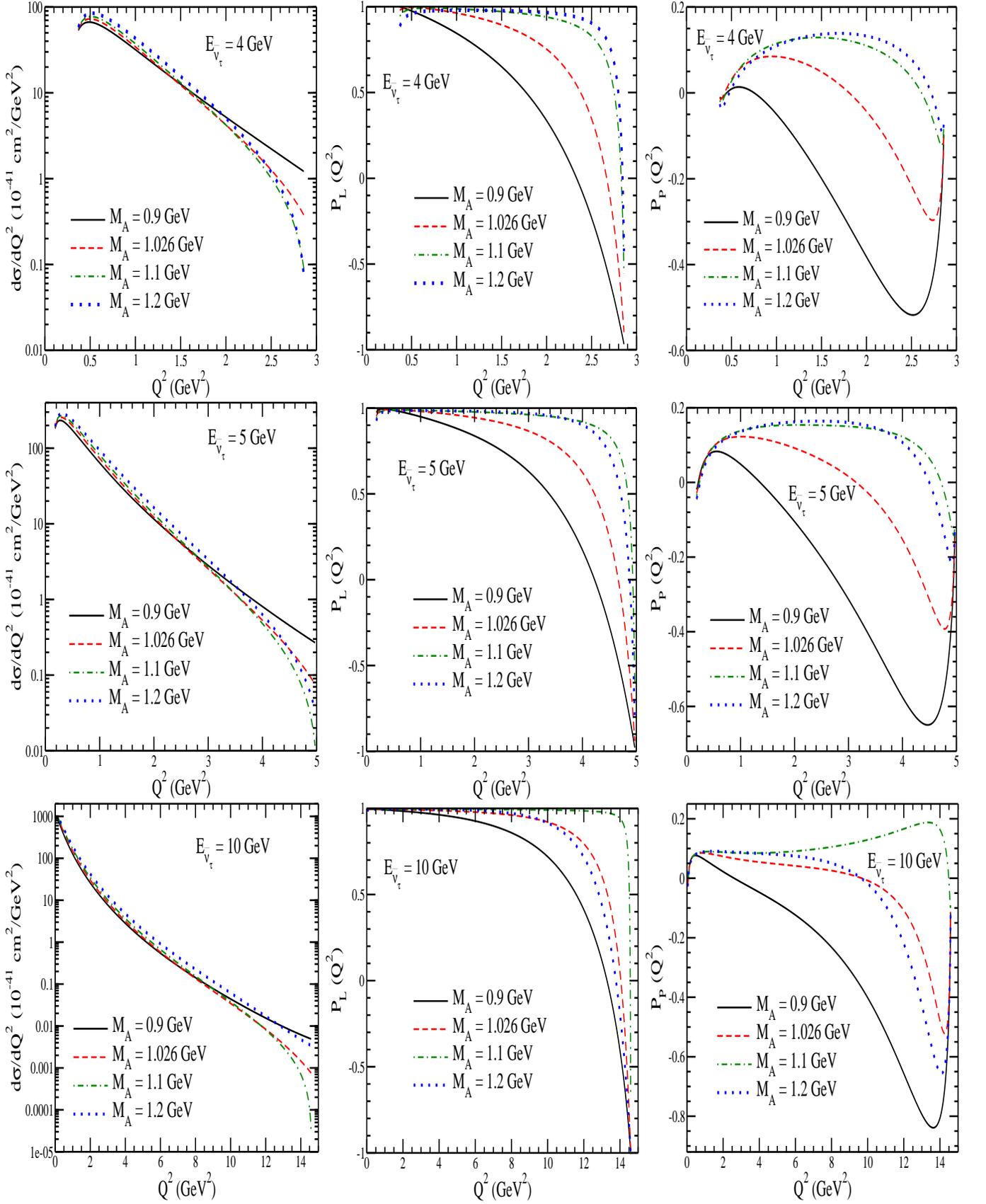

\centering
\includegraphics[height=7.5cm,width=5.9cm]{dsig_dq2_Ma_variation_enu_4GeV_neutron_polarized.eps}
\includegraphics[height=7.5cm,width=5.9cm]{Pl_q2_MA_variation_enu_4GeV.eps}
\includegraphics[height=7.5cm,width=5.9cm]{Pp_q2_MA_variation_enu_4GeV.eps}
\includegraphics[height=7.5cm,width=5.9cm]{dsig_dq2_Ma_variation_enu_5GeV_neutron_polarized.eps}
\includegraphics[height=7.5cm,width=5.9cm]{Pl_q2_MA_variation_enu_5GeV.eps}
\includegraphics[height=7.5cm,width=5.9cm]{Pp_q2_MA_variation_enu_5GeV.eps}
\includegraphics[height=7.5cm,width=5.9cm]{dsig_dq2_Ma_variation_enu_10GeV_neutron_polarized.eps} 
\includegraphics[height=7.5cm,width=5.9cm]{Pl_q2_MA_variation_enu_10GeV.eps}
\includegraphics[height=7.5cm,width=5.9cm]{Pp_q2_MA_variation_enu_10GeV.eps}
\caption{$\frac{d\sigma}{dQ^2}$~(left panel), $P_L (Q^2)$~(middle panel) and $P_P (Q^2)$~(right panel) versus $Q^2$ for the process 
$\bar{\nu}_{\tau} + p \longrightarrow \tau^+ + n$ when $\tau^+$ is polarized at $E_{\bar{\nu}_{\tau}}$ 
= 4 GeV~(upper panel), 5 GeV~(middle panel) and 10 GeV~(lower panel). The results are shown for the different values of $M_A$=0.9, 1.026, 1.1 and 1.2 GeV. Lines and points have the same meaning as in Fig.~\ref{sigma_MA_nu}.}\label{nutau_tau_pol_MA_N}
\end{figure}

\begin{figure}
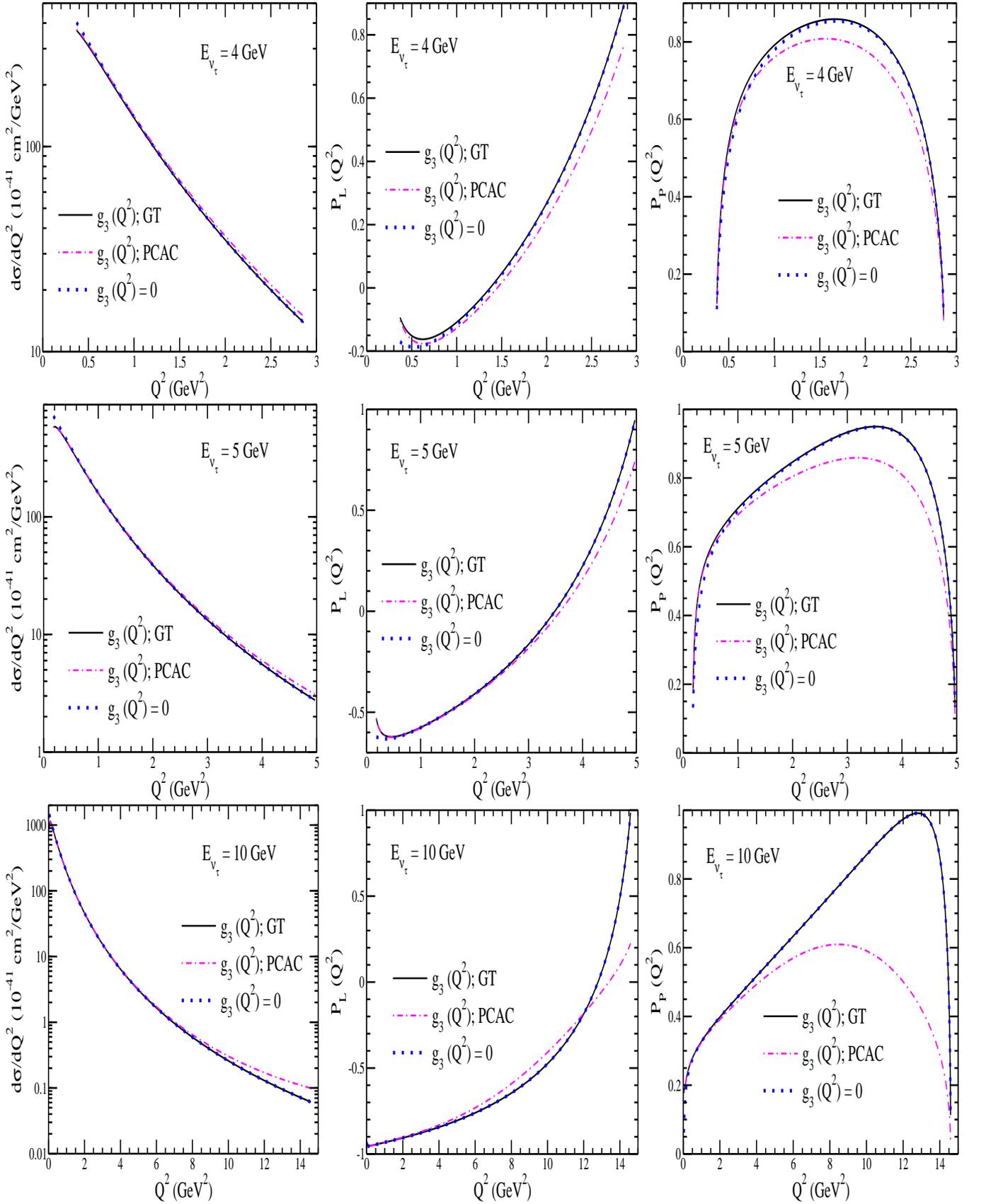

\centering
\includegraphics[height=7.5cm,width=5.9cm]{dsig_dq2_FP_variation_enu_4GeV_proton_polarized.eps}
\includegraphics[height=7.5cm,width=5.9cm]{Pl_q2_FP_variation_enu_4GeV_proton.eps}
\includegraphics[height=7.5cm,width=5.9cm]{Pp_q2_FP_variation_enu_4GeV_proton.eps}
\includegraphics[height=7.5cm,width=5.9cm]{dsig_dq2_FP_variation_enu_5GeV_proton_polarized.eps}
\includegraphics[height=7.5cm,width=5.9cm]{Pl_q2_FP_variation_enu_5GeV_proton.eps}
\includegraphics[height=7.5cm,width=5.9cm]{Pp_q2_FP_variation_enu_5GeV_proton.eps}
\includegraphics[height=7.5cm,width=5.9cm]{dsig_dq2_FP_variation_enu_10GeV_proton_polarized.eps} 
\includegraphics[height=7.5cm,width=5.9cm]{Pl_q2_FP_variation_enu_10GeV_proton.eps}
\includegraphics[height=7.5cm,width=5.9cm]{Pp_q2_FP_variation_enu_10GeV_proton.eps}
\caption{$\frac{d\sigma}{dQ^2}$~(left panel), $P_L (Q^2)$~(middle panel) and $P_P (Q^2)$~(right panel) versus $Q^2$ for the process 
$\nu_{\tau} + n \longrightarrow \tau^- + p$ when $\tau^-$ is polarized at $E_{\nu_{\tau}}$ 
= 4 GeV~(upper panel), 5 GeV~(middle panel) and 10 GeV~(lower panel) with $M_A$ = 1.026 GeV. The results are shown for the different parameterizations of the pseudoscalar form factors. Lines and points have the same meaning as in Fig.~\ref{sigma_FP_nu}.}\label{nutau_tau_pol_FP_P}
\end{figure}

In Fig.~\ref{nutau_tau_pol_MA_P}, we have presented the numerical results for the differential cross section~$\frac{d\sigma}{dQ^2}$ and the polarization components {\it viz.} $P_{L} (Q^2)$ and $P_{P}(Q^2)$ for the polarized $\tau^{-}$ in  the process $\nu_{\tau} + n \rightarrow \tau^{-} + p$ as a function of $Q^2$ at $E_{\nu_{\tau}}=$ 10 GeV by taking the different values of the axial dipole mass, $M_{A}=0.9$, 1.026, 1.1 and 1.2 GeV. For the differential cross section, we find that by increasing $M_{A}$ by 20\%, the differential cross section
 increases in the peak region by $\sim 45\%$. While for a decrease in $M_{A}$ by 10\% from the world average value, the differential cross section in the peak region decreases by $\sim 31\%$. It may be pointed out that although, there is a significant dependence of the differential cross section on $M_{A}$, the polarization observables are not much affected by the variation in $M_{A}$ at all values of $Q^2$ at a given  $E_{\nu_{\tau}}$.

\begin{figure}
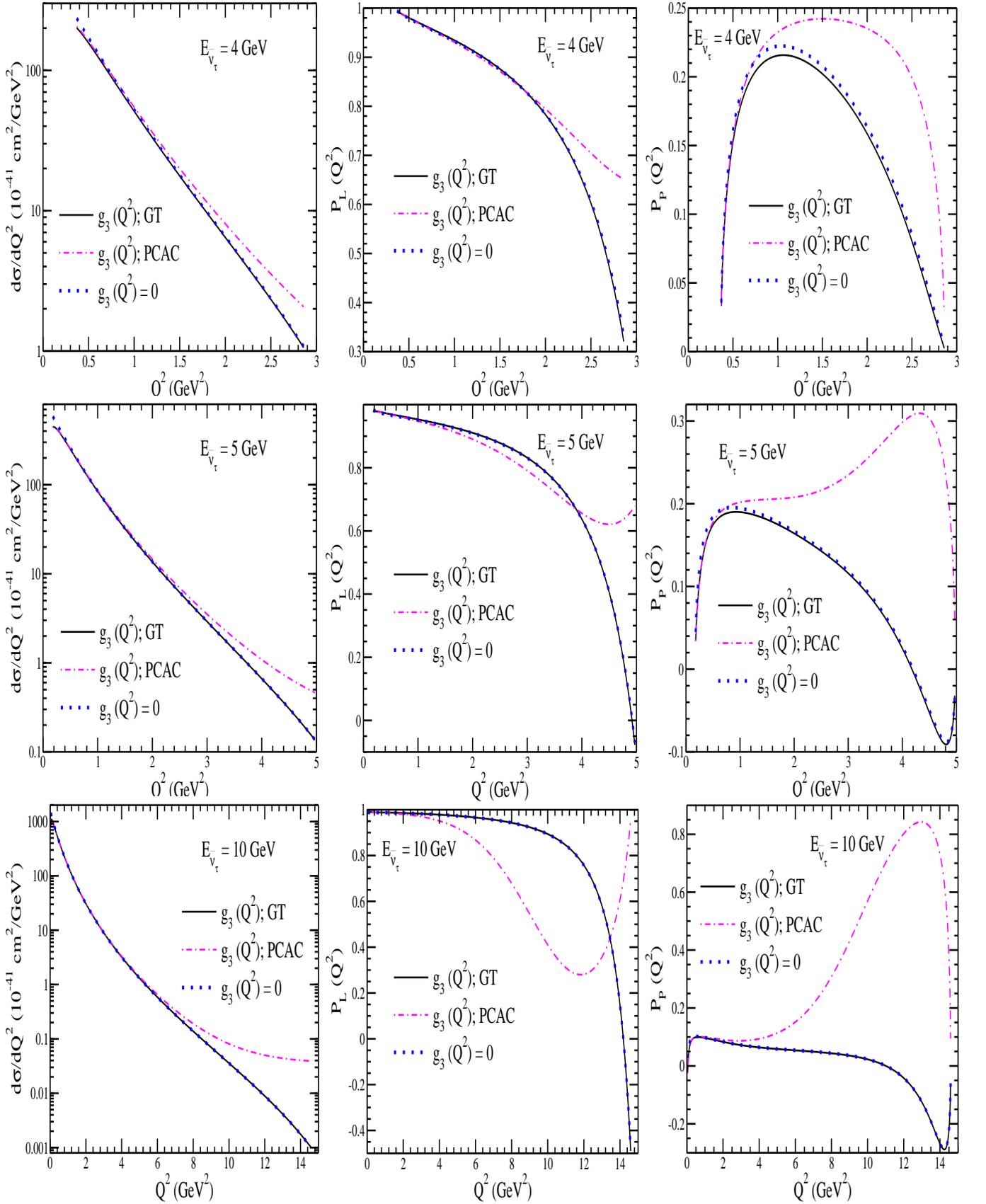

\centering
\includegraphics[height=7.5cm,width=5.9cm]{dsig_dq2_FP_variation_enu_4GeV_neutron_polarized.eps}
\includegraphics[height=7.5cm,width=5.9cm]{Pl_q2_FP_variation_enu_4GeV_neutron.eps}
\includegraphics[height=7.5cm,width=5.9cm]{Pp_q2_FP_variation_enu_4GeV_neutron.eps}
\includegraphics[height=7.5cm,width=5.9cm]{dsig_dq2_FP_variation_enu_5GeV_neutron_polarized.eps}
\includegraphics[height=7.5cm,width=5.9cm]{Pl_q2_FP_variation_enu_5GeV_neutron.eps}
\includegraphics[height=7.5cm,width=5.9cm]{Pp_q2_FP_variation_enu_5GeV_neutron.eps}
\includegraphics[height=7.5cm,width=5.9cm]{dsig_dq2_FP_variation_enu_10GeV_neutron_polarized.eps} 
\includegraphics[height=7.5cm,width=5.9cm]{Pl_q2_FP_variation_enu_10GeV_neutron.eps}
\includegraphics[height=7.5cm,width=5.9cm]{Pp_q2_FP_variation_enu_10GeV_neutron.eps}
\caption{$\frac{d\sigma}{dQ^2}$~(left panel), $P_L (Q^2)$~(middle panel) and $P_P (Q^2)$~(right panel) versus $Q^2$ for the process 
$\bar{\nu}_{\tau} + p \longrightarrow \tau^+ + n$ when $\tau^+$ is polarized at $E_{\nu_{\tau}}$ 
= 4 GeV~(upper panel), 5 GeV~(middle panel) and 10 GeV~(lower panel). Lines and points have the same meaning as in Fig.~\ref{sigma_FP_nu}.}\label{nutau_tau_pol_FP_N}
\end{figure}

In Fig.~\ref{nutau_tau_pol_MA_N}, we have studied the effect on $\frac{d\sigma}{dQ^2}$, $P_{L} (Q^{2})$ and $P_{P} (Q^{2})$ for the polarized $\tau^{+}$ produced in the process $\bar{\nu}_{\tau} + p \rightarrow \tau^{+} + n$ at 
$E_{\bar{\nu}_{\tau}} =4$, 5 and 10 GeV by varying $M_{A}$ in the range 0.9--1.2 GeV. One may notice that there is a significant 
dependence of $\frac{d\sigma}{dQ^2}$, $P_{L} (Q^{2})$ and $P_{P} (Q^{2})$ on $M_{A}$. Unlike in the case of $\nu_{\tau}$ induced quasielastic scattering~(shown in Fig.~\ref{nutau_tau_pol_MA_P}), it may be noted from the figure that in the case of $\bar{\nu}_{\tau}$ scattering, at all the values of $E_{\bar{\nu}_{\tau}}$  and at low $Q^2$, the differential scattering cross section increases with the increase in the value of $M_{A}$. We also observe significant dependence of $P_{L} (Q^2)$ and $P_{P} (Q^2)$ on $M_{A}$. Moreover, this dependence is found to be energy dependent also.

In Figs. \ref{nutau_tau_pol_FP_P} and \ref{nutau_tau_pol_FP_N}, we have presented the numerical results to depict $Q^{2}$ dependence of 
 the scattering cross section $\frac{d\sigma}{dQ^2}$ and the polarization components {\it viz.} $P_{L} (Q^2)$ and $P_{P}(Q^2)$, respectively, for the polarized $\tau^{-}$ in  the process $\nu_{\tau} + n \rightarrow \tau^{-} + p$ and $\tau^{+}$ in  the process $\bar{\nu}_{\tau} + p \rightarrow \tau^{+} + n$ at the three different values of (anti)neutrino energies {\it viz.} 4, 5 and 10 GeV using the different parameterizations of the pseudoscalar form factor given in 
 Eqs.~(\ref{g3}) and ~(\ref{fp1}). We find very small effect of the differential scattering cross section on $g_3(Q^2)$, while $P_{L}(Q^2)$ shows some dependence on $Q^2$ whereas we find a larger dependence of $P_{P}(Q^2)$ on the  pseudoscalar form factor when it is determined using the modified GT relation. Moreover, the effect is found to be larger at higher energies, for example, at $E_{\nu_\tau}=10$~GeV, considered in this work.

 \section{Summary and conclusions}\label{summary}
We have studied the quasielastic scattering of tau neutrinos and antineutrinos from nucleons induced by the 
 weak charged currents i.e. $\nu_\tau + n \rightarrow \tau^- + p$ and $\bar{\nu}_\tau + p \rightarrow \tau^+ + n$ and
 analyze the effect of using the different parameterizations of the isovector vector form factor, axial vector form factor, pseudoscalar form factor and the effect of second class currents with T invariance. The theoretical uncertainties in predicting the production cross sections and the polarization of the $\tau$ leptons due to the use of these different parameterizations have been discussed.
 The results are presented for $\sigma (E_{\nu_{\tau} (\bar{\nu}_{\tau})})$, average polarization components $\overline{P}_{L} (E_{\nu_{\tau} (\bar{\nu}_{\tau})})$ and $\overline{P}_{P} (E_{\nu_{\tau} (\bar{\nu}_{\tau})})$ for the $\tau^{\pm}$ produced in the final state.
  We have also studied the $Q^2$ dependence of the scattering cross section~($d \sigma/dQ^2$) as well as the longitudinal~($P_{L} (Q^2)$) and 
  perpendicular~($P_{P} (Q^2)$) polarization components of the $\tau$ lepton on the form factors at the different neutrino and antineutrino energies.

\vspace{2mm}

We find that:

\begin{itemize}
\item There is an appreciable difference in the results of $\sigma(E_{\nu_{\tau} (\bar{\nu}_{\tau})})$ as well as $\overline{P}_{L} (E_{\nu_{\tau} (\bar{\nu}_{\tau})})$ and ~$\overline{P}_{P} (E_{\nu_{\tau}( \bar{\nu}_{\tau})})$ when the Galster {\it et al.}~\cite{Galster:1971kv} parameterization of the vector form factors is chosen, in comparison to 
the results obtained using the other parameterizations available in the literature like that of Bradford {\it et al.}~\cite{Bradford:2006yz}, Budd {\it et al.}~\cite{Budd:2004bp}, Bosted~\cite{Bosted:1994tm}, Alberico {\it et al.}~\cite{Alberico:2008sz} and Kelly~\cite{Kelly:2004hm}.
This variation is more in the case of antineutrino induced charged current quasielastic process than in the case of neutrino induced process.

\item When the different parameterizations for the neutron electric form factor $G_{E}^{n} (Q^2)$ are chosen, then the parameterizations of BBBA05~\cite{Bradford:2006yz}, Kelly~\cite{Kelly:2004hm} and its modification~\cite{Punjabi:2015bba} give similar results, while the parameterizations of Galster~\cite{Galster:1971kv} and its modification~\cite{Platchkov:1989ch} give similar results. Moreover this variation is observed to be larger in the case of antineutrino induced charged current quasielastic process as compared to the neutrino induced process.

\item While there is a strong dependence of $M_A$ on $\sigma (E_{\nu_{\tau} (\bar{\nu}_{\tau})})$ for both neutrino as well as antineutrino induced processes, $\overline{P}_{L} (E_{\nu_{\tau}(\bar{\nu}_{\tau})})$ shows hardly any dependence for the neutrino induced process whereas for the antineutrino induced process there is a little dependence on the choice of $M_{A}$ in the low energy region, with the increase in $M_A$. The results for $\overline{P}_{P} (E_{\nu_{\tau}(\bar{\nu}_{\tau})})$ shows mild dependence on $M_A$ in the case of neutrino induced process while there is a larger dependence of $M_A$ in the case of antineutrino induced process. However, with the increase in the value of $M_A$ this difference gradually becomes smaller but nevertheless non negligible.

\item The different choices of the pseudoscalar form factor does not have much effect on the total scattering cross sections as well as on the polarization observables in the energy region of $E_{\nu_{\tau}(\bar{\nu}_{\tau})} \le 10$~GeV considered in this paper.

\item The effect of the second class current is appreciable for $g_2^R(0)=+1$~(Eq.~(\ref{g2})) in the case of $\sigma (E_{\nu_{\tau} (\bar{\nu}_{\tau})})$ both for the neutrino as 
well as antineutrino induced processes. $\overline{P}_{L} (E_{\nu_{\tau}(\bar{\nu}_{\tau})})$ for neutrino induced process shows little dependence on the second class current while for the antineutrino induced process there is significant dependence on the choice of $g_2^R(0)$
 specially for $E_{\bar{\nu}_{\tau}} <8$~GeV. $\overline{P}_{P} (E_{\nu_{\tau}(\bar{\nu}_{\tau})})$ shows strong dependence on the choice of $g_2^R(0)$ both for the neutrino as 
well as antineutrino induced processes, however, the nature of dependence is not the same for these two processes.

\item We have also studied the $Q^2$ dependence of $\frac{d\sigma}{dQ^2}$, $P_{L} (Q^{2})$ and $P_{P} (Q^{2})$ on the vector form factor, axial dipole mass and pseudoscalar form factor and the numerical results have been presented for the different (anti)neutrino energies.
\end{itemize}

 \section*{Acknowledgment}   
 M. S. A. is thankful to the Department of Science and Technology (DST), Government of India for providing 
financial assistance under Grant No. SR/MF/PS-01/2016-AMU/G.

\section*{Appendix A: Parameterization of  Sachs' electric and magnetic form factors of the nucleon}
In the following, we present the various parameterizations available in the literature for the nucleon Sachs' electric and 
magnetic form factors.
\begin{enumerate}
 \item [A.] BBBA05:\\
 The form of electric and magnetic 03 form factor given by Bradford et al.~\cite{Bradford:2006yz}~(BBBA05) is
\begin{eqnarray}\label{ge_gm_bbba01}
G_{E}^{p}(Q^2)&=&\frac{1-0.0578\tau}{1+11.1\tau+13.6\tau^2+33.0\tau^3}\nonumber\\
\frac{G_{M}^{p}(Q^2)}{\mu_{p}}&=&\frac{1+0.15\tau}{1+11.1\tau+19.6\tau^2+7.54\tau^3}\nonumber\\ 
G_{E}^{n}(Q^2)&=&\frac{1.25\tau+1.30\tau^2}{1-9.86\tau+305\tau^2-758\tau^3+802\tau^4}\nonumber\\
\frac{G_{M}^{n}(Q^2)}{\mu_{n}}&=&\frac{1+1.81\tau}{1+14.1\tau+20.7\tau^2+68.7\tau^3}, \qquad \quad \tau = \frac{Q^{2}}{4M^{2}}.
\end{eqnarray}

\item [B.] BBA03:\\
Budd {\it et al.}~\cite{Budd:2004bp}~(BBA03) parameterized electric and magnetic form factors  as
\begin{eqnarray}\label{ge_gm_bbba01}
G_{E}^{p}(Q^2)&=&\frac{1}{1 + 3.253Q^{2} + 1.422 Q^{4} + 0.08582 Q^{6} + 0.3318Q^{8} - 0.0937Q^{10} + 0.01076Q^{12}}\nonumber\\
\frac{G_{M}^{p}(Q^2)}{1+\mu_{p}}&=&\frac{1}{1 + 3.104Q^{2} + 1.428Q^{4} + 0.1112Q^{6} - 0.00698Q^{8} + 0.00037Q^{10}}\nonumber\\ 
\frac{G_{M}^{n}(Q^2)}{\mu_{n}}&=&\frac{1}{1 + 3.043Q^{2} + 0.8548Q^{4} + 0.6806Q^{6} - 0.1287Q^{8} + 0.0089Q^{12}}\nonumber\\
G_{E}^{n}(Q^2)&=&-\frac{0.942\tau}{1+4.61\tau}G_{D}(Q^{2}),
\end{eqnarray}
with $ \mu_p=1.7927\mu_N,~      \mu_n=-1.913\mu_N,~        M_V=0.84 \text{GeV}$ and $\lambda_n=5.6$ and $G_{D}(Q^{2})$ is 
parameterized as
\begin{equation}
G_{D} (Q^2) = \frac{1}{\left(1+\frac{Q^2}{M_{V}^2}\right)^{2}},
\end{equation}
with $M_{V} =0.84$ GeV and $Q^{2}$ is in units of GeV$^{2}$.
\item [C.] Galster {\it et al.}:\\
The parameterization of electric and magnetic form factors, as given by Galster {\it et al.}~\cite{Galster:1971kv}:
\begin{eqnarray}
G_E^p(Q^2)&=&G_D(Q^2) \qquad \qquad \qquad G_M^p(Q^2)=(1+\mu_p)G_{D}(Q^2) \nonumber\\
G_M^n(Q^2)&=&\mu_nG_{D}(Q^2) \qquad \qquad ~~~G_E^n(Q^2)=(\frac{Q^2}{4M^2})\mu_nG_{D}(Q^2)\xi_n \nonumber \\
\xi_n&=&\frac{1}{\left(1-\lambda_n\frac{Q^2}{4M^2}\right)}.\nonumber
\end{eqnarray}

\item [D.] Platchkov {\it et al.}~(modified Galster):\\
Platchkov {\it et al.} modified $G_{E}^{n}(Q^{2})$ of Galster's parameterization as:
\begin{equation}
 G_E^n(Q^2) = -\frac{a \mu_n \tau}{1+b\tau}G_D(Q^2),
\end{equation}
with $a$=1.51 and $b$=8.4.

\item [E.] Kelly:\\
The parameterization for $G_{E}^{p,n} (Q^2)$ and $G_{M}^{p,n} (Q^2)$ given by Kelly~\cite{Kelly:2004hm} is
\begin{eqnarray}\label{gepn_kelly}
G_{E}^{p}(Q^2)&=&\frac{1-0.24\tau}{1+10.98\tau+12.82\tau^2+ 21.97\tau^3}\nonumber\\
\frac{G_{M}^{p}(Q^2)}{\mu_{p}}&=&\frac{1+0.12\tau}{1+10.97\tau+18.86\tau^2+6.55\tau^3}\nonumber\\ 
G_{E}^{n}(Q^2)&=&\frac{1.7\tau}{1+3.3\tau} \frac{1}{\left(1 - {Q^2}/{(0.84)^2} \right)^2}\nonumber\\
\frac{G_{M}^{n}(Q^2)}{\mu_{n}}&=&\frac{1+2.33\tau}{1+14.72\tau+24.20\tau^2+84.1\tau^3}
\end{eqnarray}

\item [F.] Punjabi {\it et al.}~(modified Kelly):\\
Punjabi {\it et al.}~\cite{Punjabi:2015bba} have modified Kelly's fit \cite{Kelly:2004hm} for $G_{E}^{n}$ and $G_{E}^{p}$ by 
including the new data since the Kelly fit was done. Their best fits for $\mu_{n}G_{E}^{n}/G_{M}^{n}$ and ${\mu_{p}G_{E}^{p}}/
{G_{M}^{p}}$ are given as:
\begin{eqnarray}\label{eq:genfit}
\frac{\mu_{n}G_{E}^{n}}{G_{M}^{n}}&=&\frac{2.6316\tau}{1+4.118\sqrt{\tau}+0.29516 \tau}, \nonumber \\
    \label{eq:gepfit}
    \frac{\mu_{p}G_{E}^{p}}{G_{M}^{p}}&=&\frac{1-5.7891\tau+14.493\tau^2-3.5032\tau^3}{1-5.5839\tau+12.909\tau^2+
    0.88996\tau^3+0.5420\tau^4}. \nonumber
\end{eqnarray}

\item [G.] Alberico {\it et al.}:\\
The parameterization for $G_{E}^{p,n} (Q^2)$ and $G_{M}^{p,n} (Q^2)$ given by Alberico {\it et al.}~\cite{Alberico:2008sz} is
\begin{eqnarray}\label{ge_gm_bbba01}
G_{E}^{p}(Q^2)&=&\frac{1 - 0.19\tau}{1 + 11.12\tau + 15.16\tau^{2} + 21.25\tau^{3}}\nonumber\\
\frac{G_{M}^{p}(Q^2)}{\mu_{p}}&=&\frac{1 + 1.09\tau}{1 + 12.31\tau + 25.57\tau^{2} + 30.61\tau^{3}}\nonumber\\ 
G_{E}^{n}(Q^2)&=&\frac{1.68\tau}{1 + 3.63\tau} G_{D}(Q^{2})\nonumber\\
\frac{G_{M}^{n}(Q^2)}{\mu_{n}}&=&\frac{1+8.28\tau}{1+21.30\tau + 77\tau^{2} + 238\tau^{3}}.
\end{eqnarray}

\item [H.] Bosted:\\
The parameterization for $G_{E}^{p,n} (Q^2)$ and $G_{M}^{p,n} (Q^2)$ given by Bosted~\cite{Bosted:1994tm} is
\begin{eqnarray}\label{ge_gm_bbba01}
G_{E}^{p}(Q^2)&=&\frac{1}{1 + 0.62Q + 0.68Q^{2} + 2.80Q^{3} + 0.83Q^{4}}\nonumber\\
\frac{G_{M}^{p}(Q^2)}{1+\mu_{p}}&=&\frac{1}{1 + 0.35Q + 2.44Q^{2} + 0.5Q^{3} + 1.04Q^{4} + 0.34Q^{5}}\nonumber\\ 
G_{E}^{n}(Q^2)&=&-\mu_{n}\frac{1.25\tau}{1 + 18.3\tau} G_{D}(Q^{2})\nonumber\\
\frac{G_{M}^{n}(Q^2)}{\mu_{n}}&=&\frac{1}{1 - 1.74Q + 9.29Q^{2} - 7.63Q^{3} + 4.63Q^{4}}, 
\end{eqnarray}
where $Q=\sqrt{Q^{2}}$, is in units of GeV.
\end{enumerate}

\end{document}